\documentclass{article}
\usepackage{amsmath,amssymb, amsthm, amsbsy,bm}
\usepackage{subcaption}
\usepackage{graphicx}
\usepackage{booktabs}

\newtheorem{theorem}{Theorem}
\newtheorem{acknowledgement}[theorem]{Acknowledgement}

\begin{document}

\title{Phase field modelling of the detachment of bubbles from a solid
substrate}
\author{Carlos Uriarte \\
%EndAName
\'{A}rea de Electromagnetismo, Universidad Rey Juan Carlos, \\
Tulip\'{a}n s/n, Mostoles, 28933, Madrid, Spain \and Marco A. Fontelos \\
%EndAName
Instituto de Ciencias Matem\'{a}ticas (ICMAT, CSIC-UAM-UC3M-\\
UCM), C/ Nicol\'{a}s Cabrera 15, 28049 Madrid, Spain. \and Manuel Array\'{a}%
s \\
%EndAName
\'{A}rea de Electromagnetismo, Universidad Rey Juan Carlos, \\
Tulip\'{a}n s/n, Mostoles, 28933, Madrid, Spain}
\maketitle

\begin{abstract}
We develop and implement numerically a phase field model for the evolution
and detachment of a gas bubble resting on a solid substrate and surrounded
by a viscous liquid. The bubble has a static contact angle $\theta $ and
will be subject to gravitational forces. We compute, as a function of the
static contact angle, the cricital Bond number over which bubbles detach
from the substrate. Then, we perform similar studies for bubble resting on
inclined substrates and bubbles under the action of an external flow. We
provide approximate formulas for the critical Bond number under all these
circumstances. Our method is also able to resolve the pinchoff of the bubble
and the possible appearence of satellites.
\end{abstract}

\section{\protect\bigskip Introduction}

The nucleation, growth and detachment of gas bubbles surrounded by a viscous
liquid is a classical problem in fluid mechanics. It appears in contexts as
diverse as boiling of liquids (cf. \cite{Z}), cavitation (cf. \cite{FA}),
bubble creation and transport in microchannels (cf. \cite{JO}) or
electrochemistry (cf. \cite{ZLF}). In the context of electrochemistry, where
chemical reactions at an electrode can lead to the production of molecules
in a gas phase so that bubbles nucleate and grow attached to it. When the
bubble's volume is sufficiently large, the bubble can detach and carry the
produced gas with it. This is the case, for instance, of Hydrogen produced
by electrolysis of water which is a process of enormous industrial
importance in connection to the storage and transport of energy produced by
renewable means such as solar and wind energies (see for instance \cite{YL}%
). It is then natural to search for methods and techniques to optimize the
processes of electrolysis in order to increase energy production (cf. \cite%
{YY}).

The life cycle of a bubble at a gas-evolving electrode begins with its
nucleation at a suitable site of the electrode surface. The bubble grows by
taking up dissolved gas that reaches its surface by diffusion, and detaches
from the electrode when the buoyancy force, aided by hydrodynamic forces if
the liquid ows around the electrode, overcomes the surface tension and
electric forces that keep the bubble attached to the electrode surface. The
detached bubble then drifts in the liquid until it reaches the surface where
the gas is collected. Coalescence of bubbles may occur before and after
detachment. The main difficulty when modelling a bubble and its dettachment
is the presence of moving interfaces (the surface of the bubble) separating
different media. This forces the implementation of suitable boundary
conditions at a surface that evolves in time. A useful approach to
simulation of problems involving moving interfaces separating two different
phases is by means of a so-called phase field (see \cite{C}, \cite{DF} for a
general description of the method and \cite{J}, \cite{VCB}, \cite{TQ}, \cite%
{EFG}, \cite{FGK} for its application for fluid mechanical problems). The
phase field approach replaces (in a way to be described below) the sharp
interface by a so-called diffuse interface across which a phase field
function changes smoothly. This removes the difficulty of numerically
tracking the interface which is replaced by a suitable level surface of the
phase field (cf. \cite{DF}). In addition, one can handle topological changes
such as those produced in the detachment or coalescence of bubbles (see \cite%
{ZLF}).

\begin{figure}[t]
\includegraphics[width=1.0\textwidth]{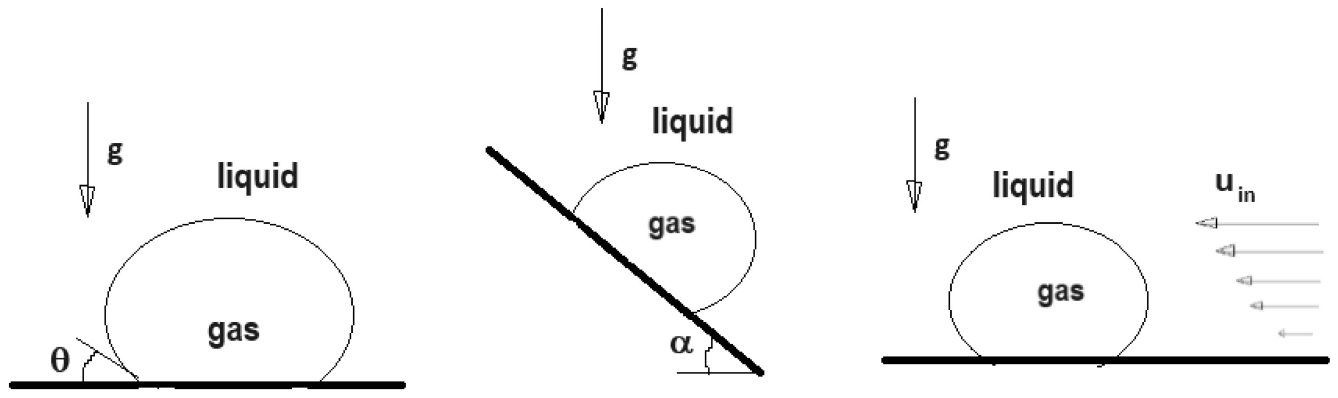}
\caption{Sketch of physical settings: bubble under gravity (left), over a
inclined substrabe (center) and under external flow \ and gravity (right). $%
\protect\theta $ is the contact angle and $\protect\alpha $ the inclination
angle.}
\label{fig1}
\end{figure}

In this article we implement and use a phase field model for the growth and
detachment of gas bubbles from a solid substrate under various circumstances
represented in figure \ref{fig1}. Section 2 will be devoted to the deduction
of a suitable phase field model coupling with Navier-Stokes equations and
being able to resolve a predetermined contact angle condition. In section 3
we will use this model to study the evolution and stability of bubbles of
different volumes (more precisely, for different Bond numbers). In section 4
we consider the sitiation of an inclined substrate on bubble detachment.
Finally, in section 4 we impose an external flow and study its influence in
bubble detachment. We will develop a phase plane for bubble detachment
conditions as a function of both volume and imposed external flow and will
do it for multiple contact angles.

\section{\protect\bigskip Phase field modelling}

One of the most effective methods to study multiphase flows is the so-called
phase field method. The main idea is to replace sharp interfaces by
``diffuse" interfaces where a phase field function $\phi (\mathbf{x},t)$
experiences sharp transitions (across the diffuse interface with a thickness
of order $\varepsilon $ sufficiently small) between to limiting values (say $%
\phi =1$ and $\phi =-1$, for instance) corresponding to two different
fluids. The introduction of diffuse interfaces has, as one of its main
advantages, the property that topological changes in the different fluid
domains can be dealt with easily. One needs to provide a suitable PDE to the
phase field $\phi (\mathbf{x},t)$ and couple to other fluid variables in a
proper way. This has been studied in many papers (cf. \cite{J}, \cite{VCB},
\cite{TQ}, \cite{EFG}, \cite{FGK} for instance). For the phase field
function, the suitable equation is a fourth order PDE called the
Cahn-Hilliard equation (firstly introduced in \cite{CH}, see also \cite{AC})
including a convection term with a velocity $\mathbf{v}(\mathbf{x},t)$ which
is the fluid velocity:

\begin{equation}
\frac{\partial \phi }{\partial t}+\mathbf{v}\cdot \nabla \phi =\nabla \cdot
\left( M\nabla \psi \right),  \label{b1}
\end{equation}%
with%
\begin{equation}
\psi =-\varepsilon \Delta \phi +\frac{1}{\varepsilon }W^{\prime }(\phi ),
\label{b2}
\end{equation}%
where $\psi $ is the so-called chemical potential.

In (\ref{b1}) $M$ is a ``mobility" factor and in (\ref{b2}) $W(\phi )$ is a
phase-field potential with the property of having two local minima at values
of $\phi $ corrresponding to the two phases. In particular we will take $%
W(\phi )=\phi ^{2}(1-\phi ^{2})$.

Concerning Navier-Stokes system, we have
\begin{equation*}
\frac{\partial (\rho \left( \phi \right) \mathbf{v)}}{\partial t}+\rho
\left( \phi \right) \mathbf{v}\cdot \nabla \mathbf{~v}-\nabla \cdot \mathbf{S%
}=-\nabla p+\mu \nabla \phi -\rho \left( \phi \right) g\mathbf{e}_{z},
\end{equation*}%
where the viscous stress tensor is given by%
\begin{equation*}
\mathbf{S}=\frac{\mu \left( \phi \right) }{2}\left( \nabla \mathbf{v}+\nabla
\mathbf{v}^{T}\right),
\end{equation*}%
and the material parameters $\rho \left( \phi \right) $ and $\mu \left( \phi
\right) $ interpolate between the fluids densities and viscosities
respectively:

\begin{eqnarray*}
\rho \left( \phi \right)  &=&\rho _{1}\phi +\rho _{2}(1-\phi ), \\
\mu \left( \phi \right)  &=&\mu _{1}\phi +\mu _{2}(1-\phi ).
\end{eqnarray*}%
We will also assume fluid incompressibility:%
\begin{equation*}
\nabla \cdot \mathbf{v}=0,
\end{equation*}%
and for the phase field the boundary condition \
\begin{equation}
\frac{\partial \psi }{\partial n}=0  \label{b3}.
\end{equation}%
The condition (\ref{b3}) is not sufficient for a fourth order equation such
as (\ref{b1}) and we need another boundary condition. Following \cite{EFG}
and \cite{TQ}, we will impose the condition
\begin{equation}
\sigma _{0\text{ }}\varepsilon \frac{\partial \phi }{\partial n}=\sigma
_{fs}^{\prime }(\phi ),  \label{b4}
\end{equation}%
where $\sigma _{fs}^{\prime }(\phi )$ interpolates between the liquid/solid
interfacial energy $\sigma _{LS}$ and the gas/solid interfacial energy $%
\sigma _{GS}$:%
\begin{equation*}
\sigma _{fs}(\phi )=\frac{\sigma _{GS}+\sigma _{LS}}{2}+\frac{\sigma
_{GS}-\sigma _{LS}}{2}\sin (\frac{\pi \phi }{2}),
\end{equation*}%
and $\sigma _{0}$ is proportional to the liquid/gas interfacial energy $%
\sigma _{LG}$:%
\begin{equation*}
\sigma _{0}=\frac{3\sqrt{2}}{8}\sigma _{LG} \,.
\end{equation*}%
In this way, the condition (\ref{b4}) imposes, in the limit $\varepsilon
\rightarrow 0$, a liquid/gas contact (or Young's) angle $\theta _{Y}$ such
that%
\begin{equation*}
\cos \theta _{Y}=\frac{\sigma _{GS}-\sigma _{LS}}{\sigma _{LG}}\,.
\end{equation*}

The phase field model described in this section has been implemented
numerically by means of a finite element method implemented in COMSOL. The
initial bubble has a radius $r=1$ (dimensionless units) and lied inside a
computational rectangular box of size $10 \times 10 \times 20$ dimensionless units. The
computational domain has been discretized by using approximately $%
N=2.3\times 10^{6}$ elements.

\section{Transient evolution and stability/instability}

In this section we will study the evolution and detachment (or not) of a
bubble on a solid substrate and inside a viscous liquid. We will base our
theoretical discussion on the sharp interface description and present
numerical results for the diffuse interface approximation.

\begin{equation*}
\frac{\partial (\rho _{i}\mathbf{v)}}{\partial t}+\rho _{i}\mathbf{v}\cdot
\nabla \mathbf{~v}-\nabla \cdot \mathbf{S}_{i}=-\nabla p-\rho _{i}g\mathbf{e}%
_{z},
\end{equation*}%
with a viscous stress%
\begin{equation*}
\mathbf{S}_{i}=\frac{\mu _{i}}{2}\left( \nabla \mathbf{v}+\nabla \mathbf{v}%
^{T}\right),
\end{equation*}%
and the balance of force condition at the interface
\begin{equation*}
\left[ \left( -p\mathbf{I}+\mathbf{S}_{i}\right) \cdot \mathbf{n}\right]
=\sigma \kappa \mathbf{n}\, ,
\end{equation*}%
where $\kappa (\mathbf{x},t)$ is the mean curvature at any point $\mathbf{x}$
at the surface of the bubble in contact with the liquid and $\sigma$ represents the surface tension between the liquid and gas interfaces. We consider, as
initial data, a bubble consisting of a half-sphere with radius $R$. Note
that $R$ may be related to the volume $V$ of the bubble by means of the
relation%
\begin{equation*}
V=\frac{2}{3}\pi R^{3}.
\end{equation*}

The introduction of a characteristic velocity $U$ defined from the following
balance between inertial and gravitational forces%
\begin{equation*}
\rho _{1}\frac{U^{2}}{R}=\rho _{1}g\, ,
\end{equation*}%
allows to write the system in terms of the following dimensionless numbers
\begin{equation*}
Bo=\frac{gR^{2}\rho _{1}}{\sigma },\ \ Re=\frac{\rho UR}{\mu },
\end{equation*}%
where $Bo$ is the Bond number and $Re$ the Reynolds number, as well
as the density and viscosity ratios%
\begin{equation*}
\gamma =\frac{\rho _{2}}{\rho _{1}},\ \delta =\frac{\mu _{2}}{\mu _{1}}.
\end{equation*}%
Hence, the velocity field satisfies%
\begin{eqnarray*}
\frac{\partial \mathbf{v}}{\partial t}+\mathbf{v}\cdot \nabla \mathbf{~v}-%
\frac{1}{Re}\Delta \mathbf{v} &=&-\nabla p-Bo\, \mathbf{e}_{z}, \ \ \text{%
in the liquid phase}, \\
\frac{\partial \mathbf{v}}{\partial t}+\mathbf{v}\cdot \nabla \mathbf{~v}-%
\frac{\delta }{Re}\Delta \mathbf{v} &=&-\nabla p-\gamma Bo\,\mathbf{e}%
_{z}, \ \text{in the gas phase}.
\end{eqnarray*}%
We are going to compare the energy of a detached bubble with the energy of a
bubble attached to a solid substrate. No external flow is considered. There
are two energies to consider: 1) Interfatial energies%
\begin{eqnarray*}
E_{s} &=&\sigma _{LG}A_{LG}+\sigma _{LS}A_{LS}+\sigma _{GS}A_{GS} \\
&=&\sigma _{LG}A_{LG}+\sigma _{LS}\left( A_{S}-A_{GS}\right) +\sigma
_{GS}A_{GS} \\
&=&C+\sigma _{LG}A_{LG}+\left( \sigma _{GS}-\sigma _{LS}\right) A_{GS}\, ,
\end{eqnarray*}%
where $A_{LG}$, $A_{LS}$ and $A_{GS}$ are the liquid/gas, liquid/solid and
gas/solid interfacial areas respectively and $\sigma _{LG}$, $\sigma _{LS}$
and $\sigma _{GS}$ the corresponding surface tension coefficients. The
constant $C=\sigma _{LS}A_{S}$ is the interfacial energy when the liquid
wets the whole solid substrate (with area $A_{S}$). The energy with the
bubble detached is%
\begin{equation*}
E_{s}^{d}=C+\sigma _{LG}A_{LG}^{d}.
\end{equation*}%
We can approximate, for small contact angles (measured in the liquid phase),%
\begin{equation*}
A_{LG}^{d}\simeq A_{LG}+A_{GS}.
\end{equation*}%
Hence, since%
\begin{equation*}
\frac{\sigma _{GS}-\sigma _{LS}}{\sigma _{LG}}=\cos \theta \, ,
\end{equation*}%
we have%
\begin{equation*}
\Delta E_{s}=\sigma _{LG}(1-\cos \theta )A_{GS}\, .
\end{equation*}

We compute now the potential energy. Concerning the detached bubble (assumed
spherical and of radius $r$), it is%
\begin{equation*}
E_{p}^{d}=(\rho _{L}-\rho _{G})gVr\, ,
\end{equation*}%
while for the attached bubble%
\begin{equation*}
E_{p}=(\rho _{L}-\rho _{G})gV(r-\delta h),
\end{equation*}%
where $\delta h$ is the difference in height between a spherical detached
bubble and an attached spherical cap with contact angle $\theta $. The
difference in potential energy is, up to $O(\delta h^{2})$ error,
\begin{equation}
\Delta E_{p}=(\rho _{L}-\rho _{G})gV\delta h  \label{c1}.
\end{equation}%
For small contact angle $\theta $, the liquid/gas interface can be
approximated by the parabola%
\begin{equation*}
y=\frac{\kappa }{2}x^{2},
\end{equation*}%
where $\kappa $ is the mean curvature (inverse of the radius of curvature),
and the volume by%
\begin{equation*}
\int_{0}^{\delta h}\pi \frac{2y}{\kappa }dy=\frac{\pi }{\kappa }(\delta
h)^{2}=\frac{1}{2}A_{GS}\delta h.
\end{equation*}%
Next, since%
\begin{equation*}
\delta h\sim \frac{1}{2}\kappa r^{2}\sin ^{2}\theta,
\end{equation*}%
we conclude%
\begin{equation}
A_{GS}=\frac{2}{\kappa }\pi \delta h\sim \pi r^{2}\sin ^{2}\theta.  \label{c3}
\end{equation}%
The difference in interfacial energies between atached and detached bubble
is then
\begin{equation}
\Delta E_{p}=(\rho _{L}-\rho _{G})gV\frac{\kappa }{2\pi }A_{GS}=(\rho
_{L}-\rho _{G})g\frac{2}{3}r^{2}A_{GS}.  \label{c2}
\end{equation}%
By comparing (\ref{c1}), (\ref{c2}) and using (\ref{c3}) we conclude the
following condition for bubble detachment:%
\begin{equation*}
\frac{(\rho _{L}-\rho _{G})gr^{2}}{\sigma _{LG}}=\frac{3}{2}(1-\cos \theta ),
\end{equation*}%
where $r$ is the radius of the full sphere. In terms of the radius $R$ of
the initial half-sphere, since $r=2^{-\frac{1}{3}}R$, the critical Bond
number would be%
\begin{equation}
Bo=\frac{(\rho _{L}-\rho _{G})gR^{2}}{\sigma _{LG}}=\frac{3}{2^{\frac{1}{3}}}%
(1-\cos \theta ).  \label{c4}
\end{equation}%
Notice that the discussion above is restricted to small values of $\theta $
so that the critical value of Bond for detachment is $O(\theta ^{2})$. We
will test this result numerically and, more interestingly, will find out
that the parabolic dependence of the critical $Bo$ on $\theta $ extends to
all values of $\theta $. We have simulated the evolution of a bubble that
initially is a sphere section with the phase field method described
previously. After a quick relaxatation towards a bubble with the
corresponding contact angle (that we establish a priori), the bubble evolves
towards a stationary configuration for small $Bo$. This stationary
configuration changes as $Bo$ increases until $Bo$ reaches a critical value
and the bubble detaches.

\begin{figure}[t]
\center
\includegraphics[width=0.7\textwidth]{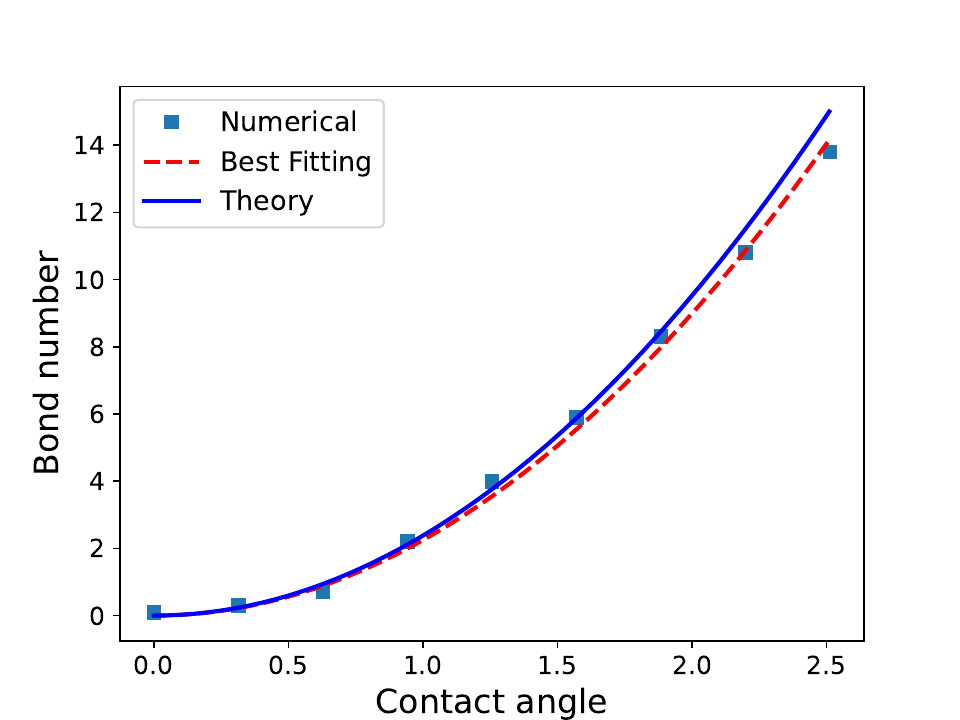}
\caption{Bond critical number versus contact angle. Squares are numerical
simulations, blue is theory and discontinuos line is the parabolic fitting
of the numerical simulations $Bo=l\protect\theta^2$, being $l_{fit}=2.25$
and $l_{th}=2.38$.}
\label{fig2}
\end{figure}

In figure \ref{fig2} we represent the critical Bond number as a function of
the contact angle. The numerical simulations are fitted using a parabolic
expression relating the critical Bond number with the contact angle, $Bo= l
\theta^2$, being $l=2.25$. The theoretical curve is derived as an asymptotic
expression from (\ref{c4}), $Bo= l_{th} \theta^2$, being $%
l_{th}=3/2^{1/3}=2.38$. We find that the agreement is satisfactory given the
simple arguments based on energy balance.

\begin{figure}[t]
\centering
\begin{subfigure}{0.4\textwidth}
    \includegraphics[width=\textwidth]{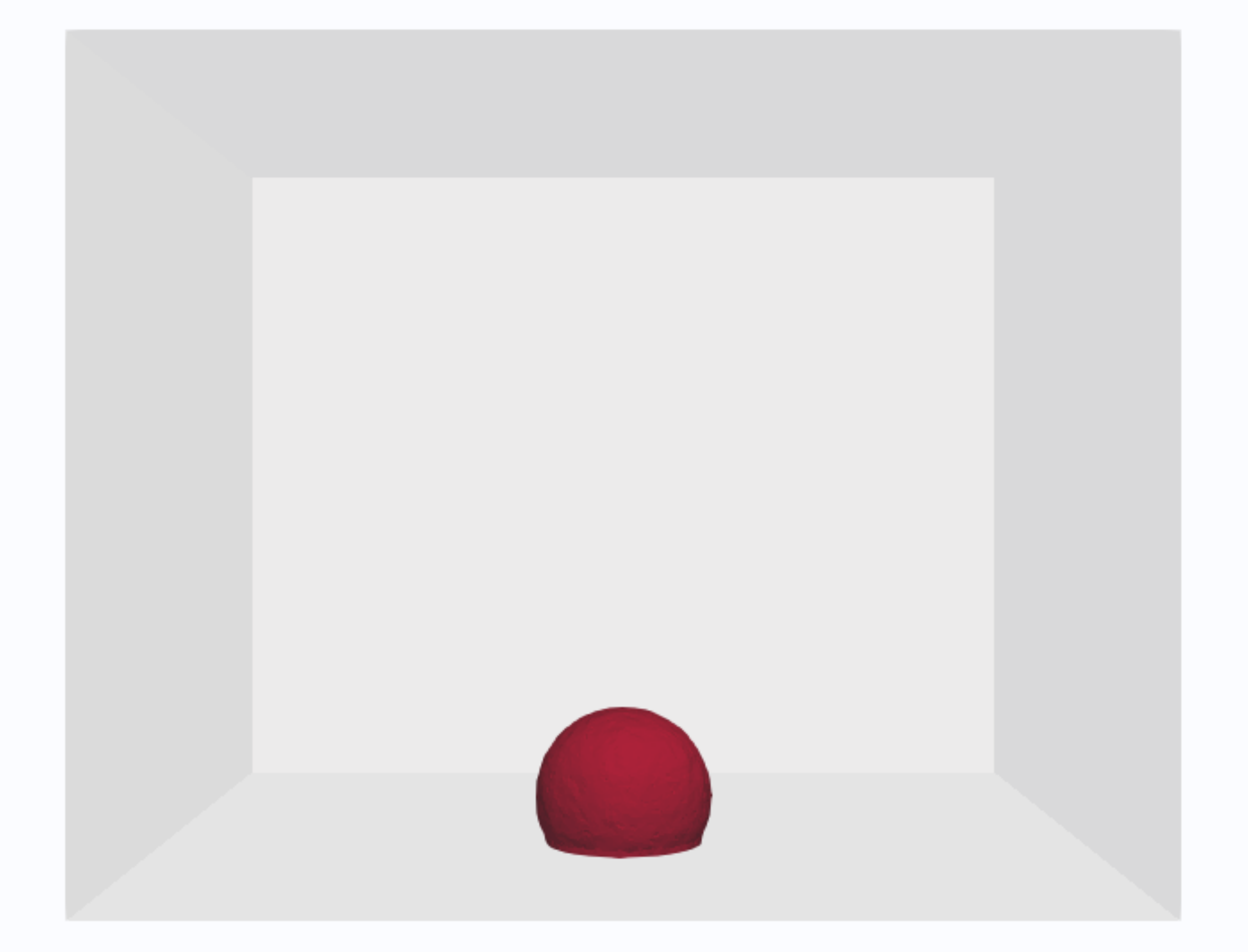}
    \caption{Initial state, $t=0$.}
    \label{fig:first_s}
\end{subfigure}
\hfill
\begin{subfigure}{0.4\textwidth}
    \includegraphics[width=\textwidth]{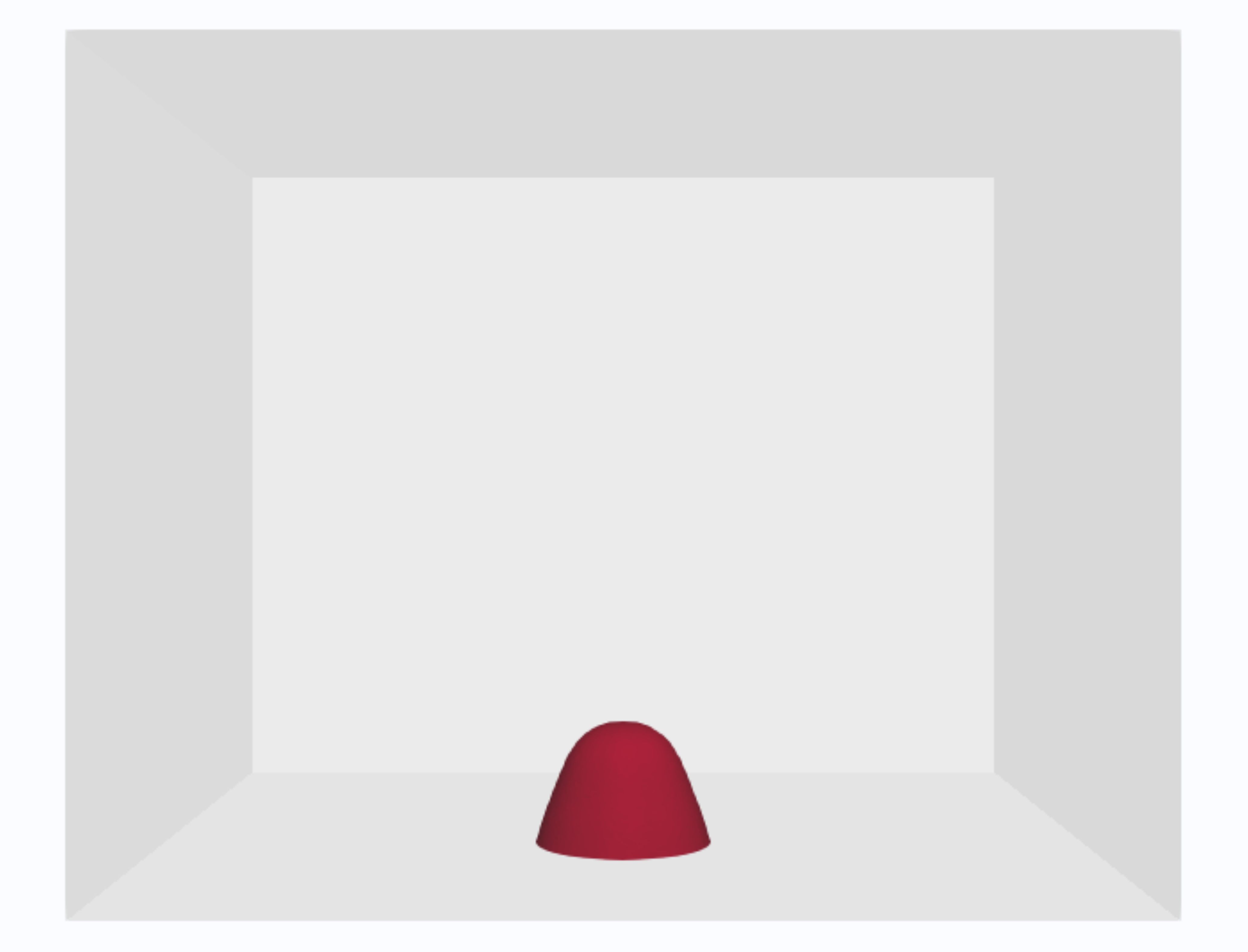}
    \caption{Time evolution, $t=0.1$}
    \label{fig:second_s}
\end{subfigure}
\hfill
\begin{subfigure}{0.4\textwidth}
    \includegraphics[width=\textwidth]{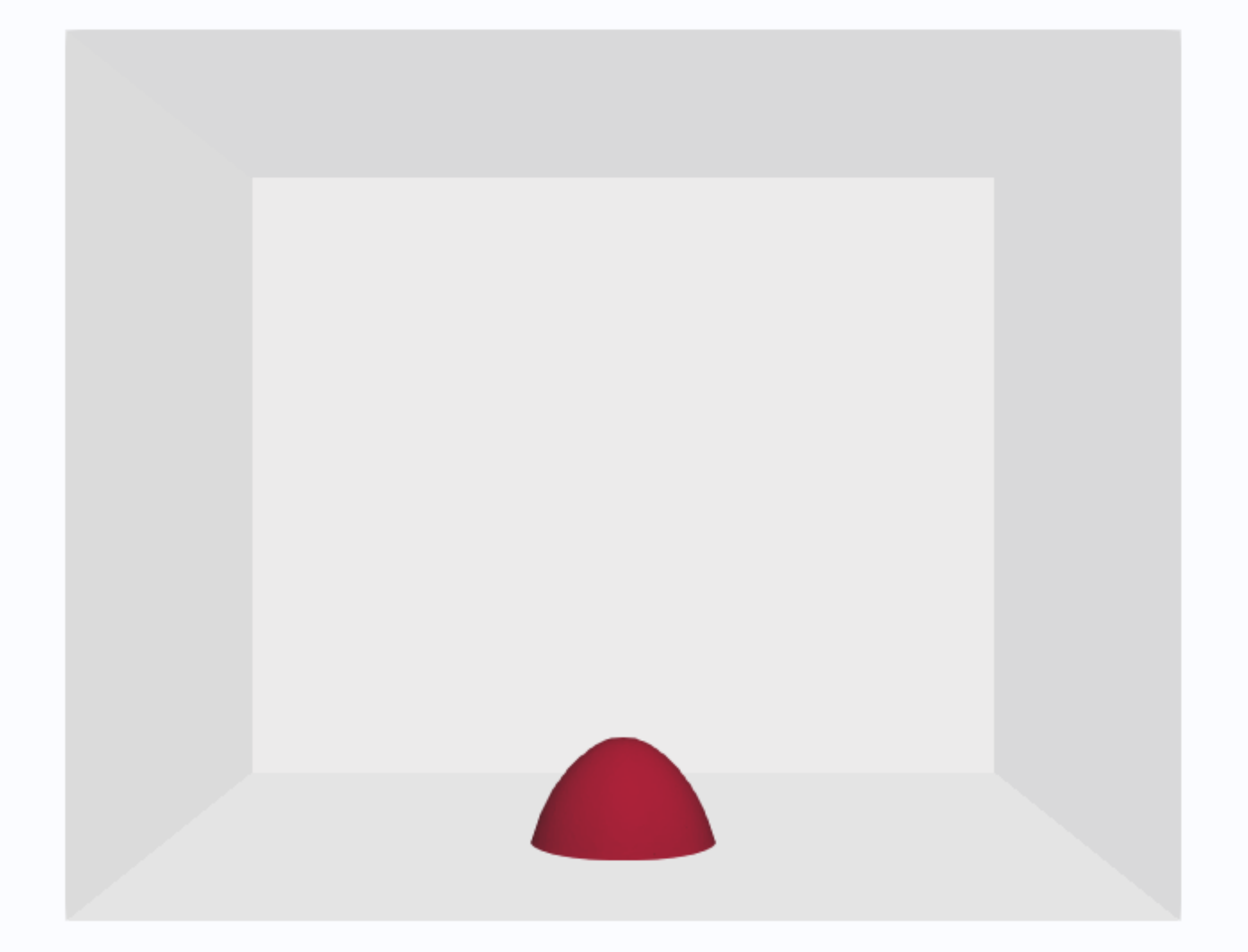}
    \caption{The relaxation process at, $t=0.2$.}
    \label{fig:third_s}
\end{subfigure}
\hfill
\begin{subfigure}{0.4\textwidth}
    \includegraphics[width=\textwidth]{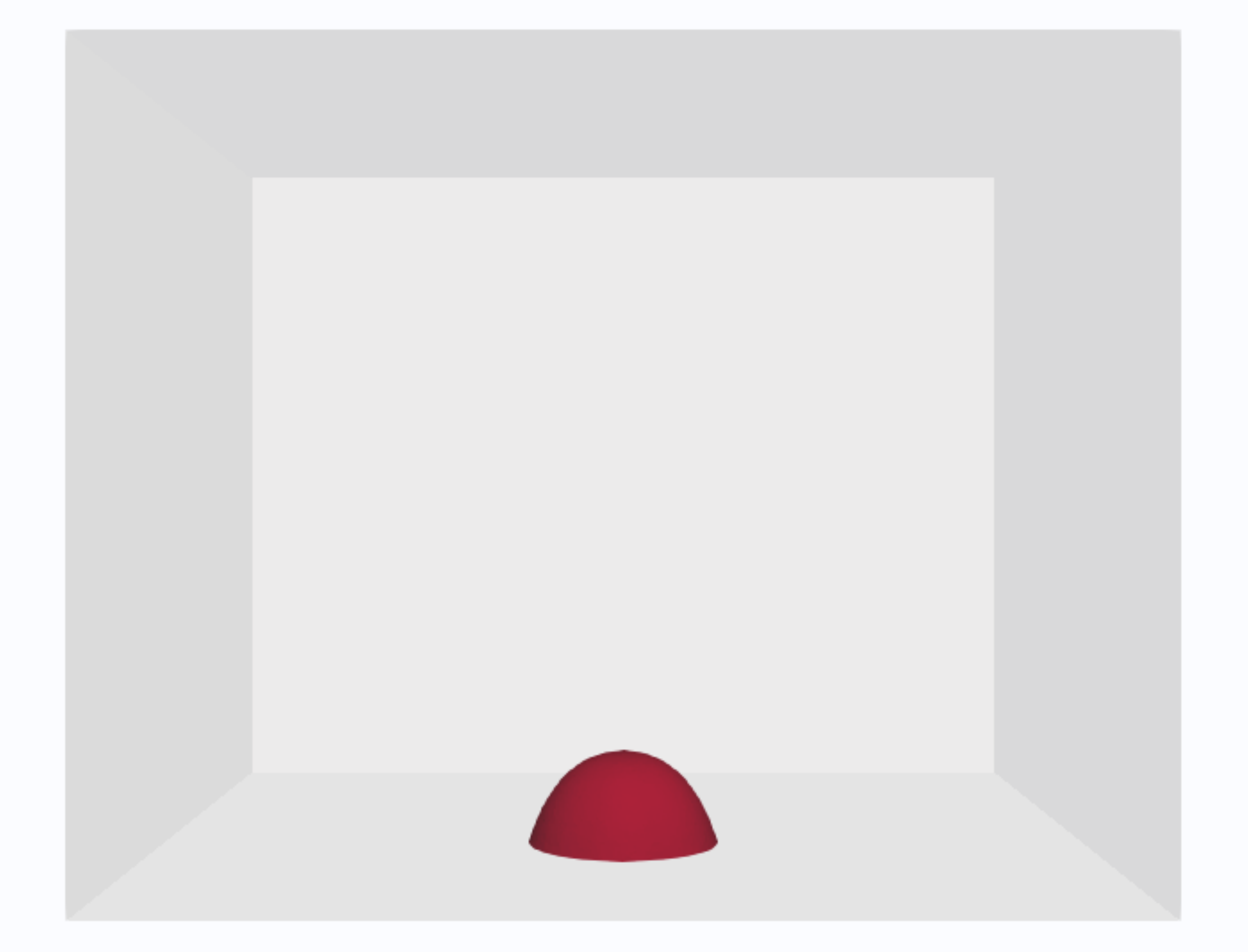}
    \caption{Towards equilibrium, $t=0.3$.}
    \label{fig:fourth_s}
\end{subfigure}
\hfill
\begin{subfigure}{0.4\textwidth}
    \includegraphics[width=\textwidth]{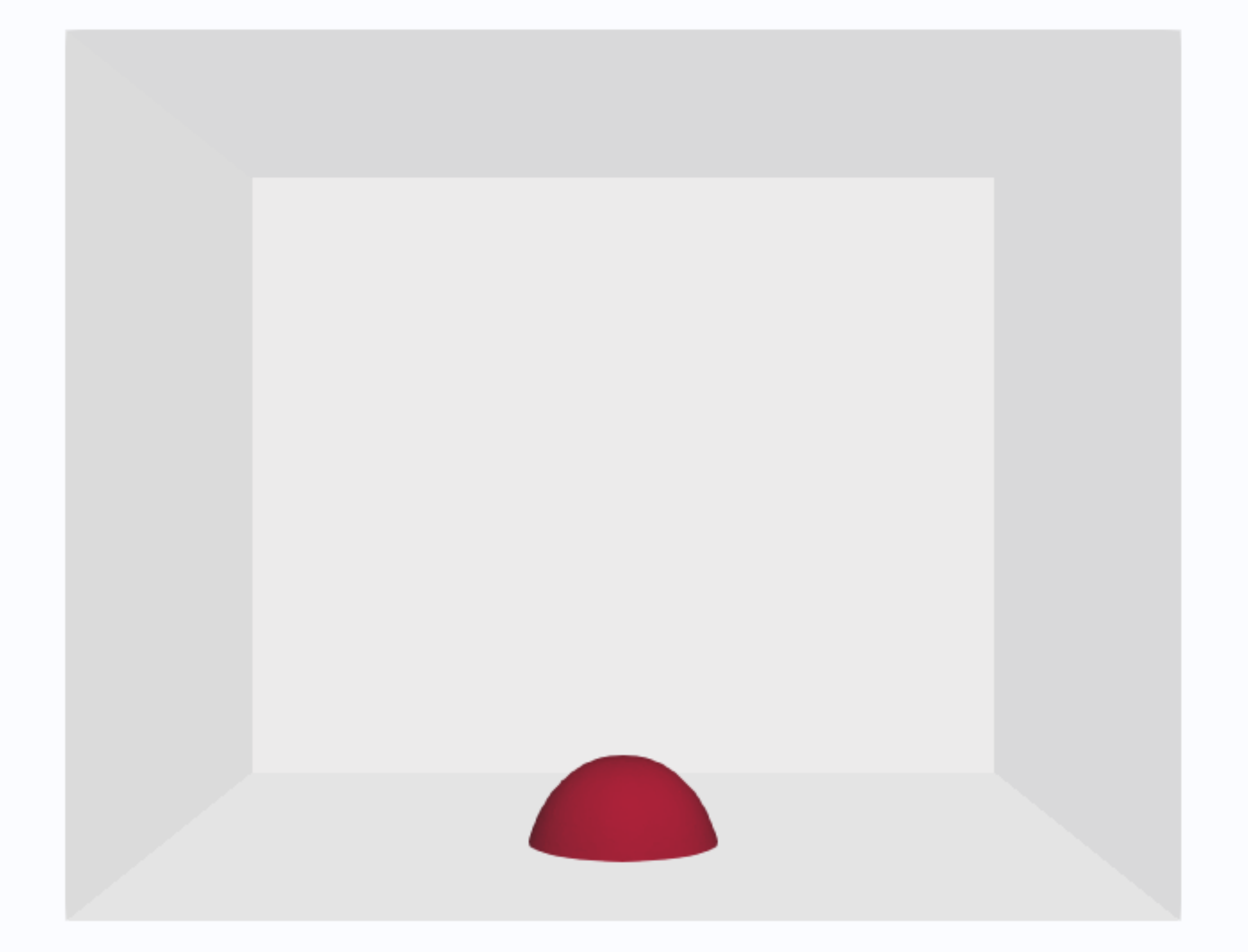}
    \caption{Reaching equilibrium state, $t=0.4$.}
    \label{fig:fifth_s}
\end{subfigure}
\hfill
\begin{subfigure}{0.4\textwidth}
    \includegraphics[width=\textwidth]{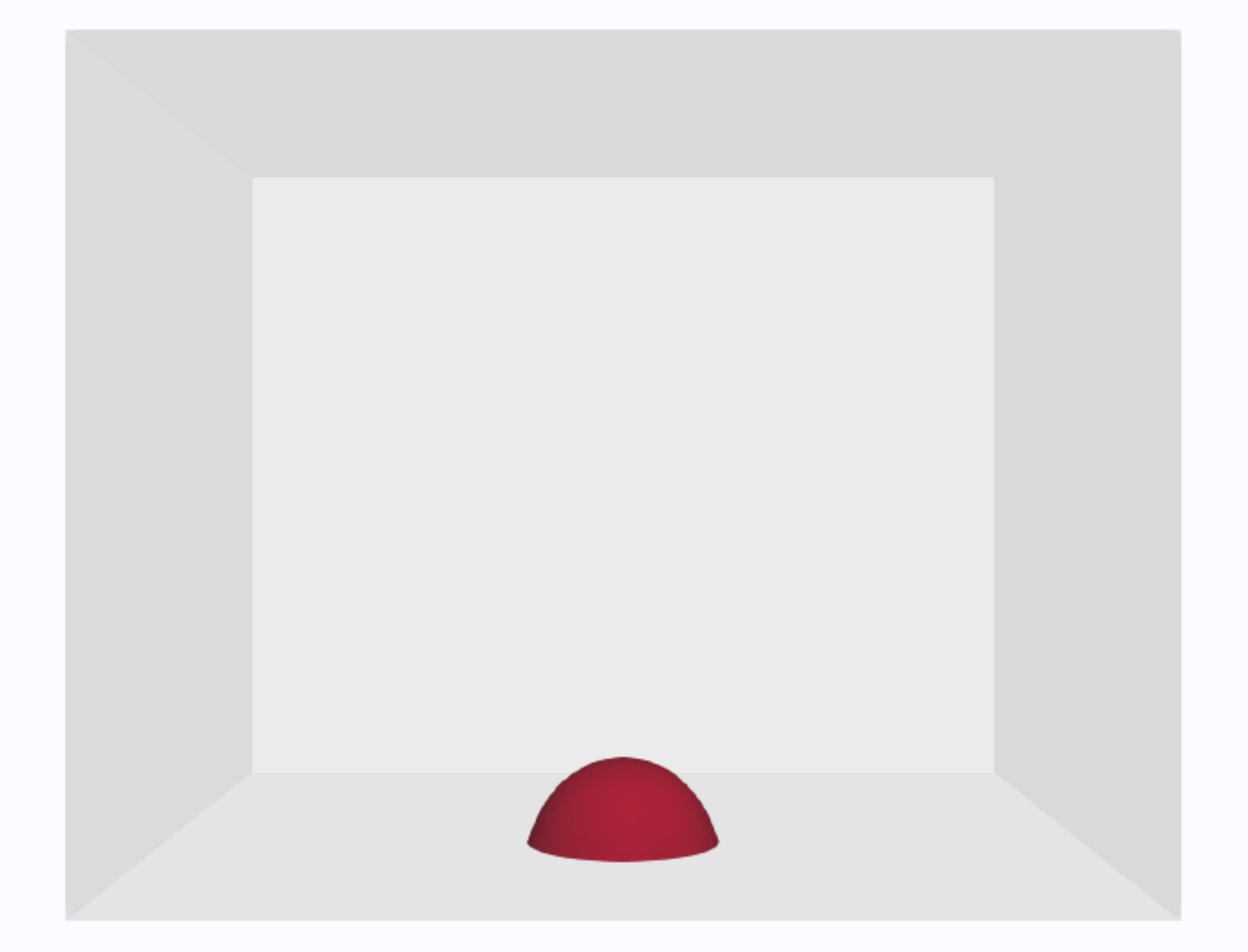}
    \caption{Equilibrium state, $t=0.5$.}
    \label{fig:sixth_s}
\end{subfigure}
\caption{Numerical simulations of the evolution of a bubble with initial
contact angle $3\protect\pi/5$ and Bond number 4 (time in dimensionless
units), below criticality.}
\label{fig_lowbond}
\end{figure}

In figure \ref{fig_lowbond} we plot some snapshots of the evolution of a
bubble. We have simulated the case $Bo=4$, and contact angle $\theta =3\pi/5$%
. At those values, we are below the critical curve so we are through the
process of relaxation to equillibrium without detachment.

\begin{figure}[t]
\centering
\begin{subfigure}{0.4\textwidth}
    \includegraphics[width=\textwidth]{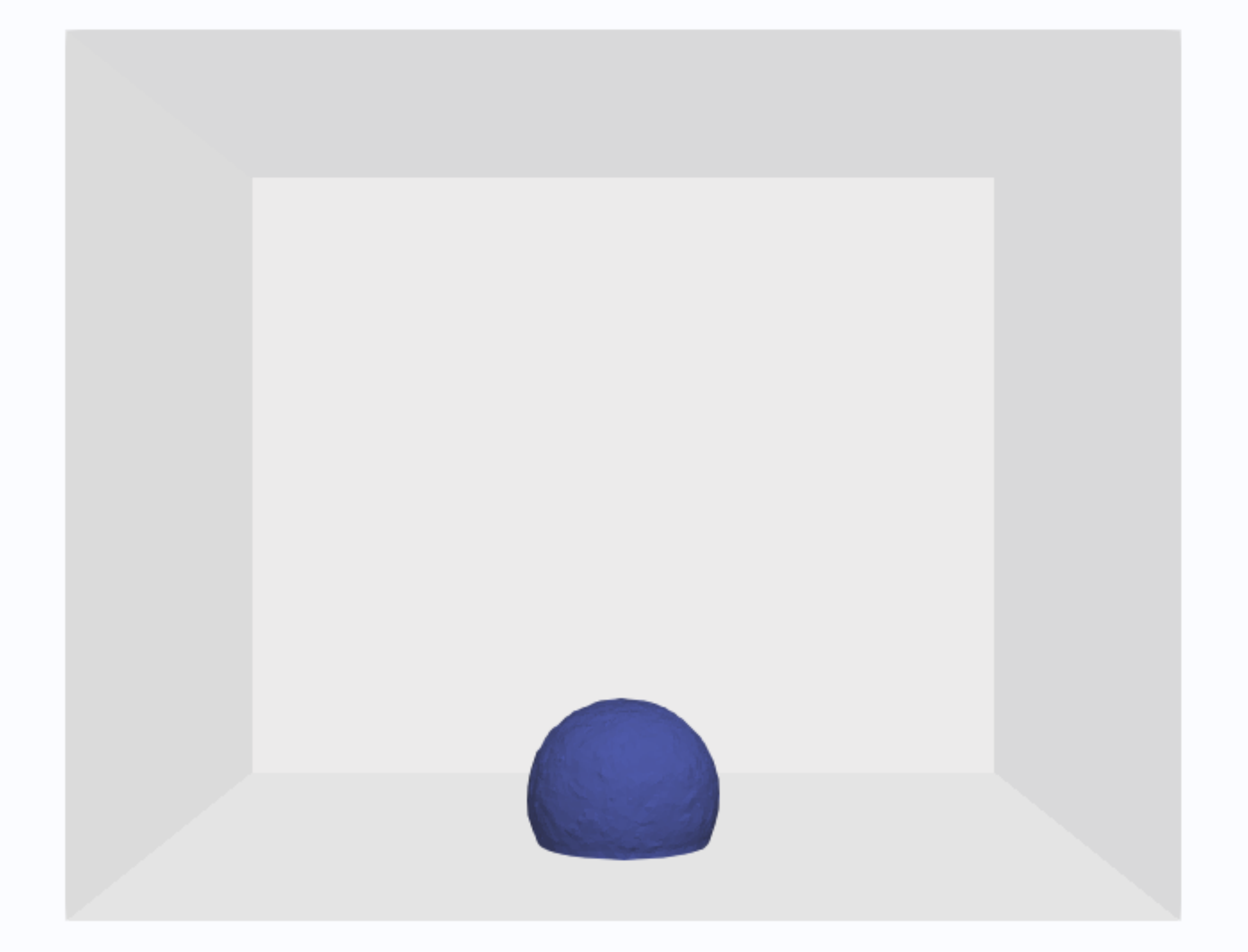}
    \caption{Initial state, $t=0$.}
    \label{fig:first}
\end{subfigure}
\hfill
\begin{subfigure}{0.4\textwidth}
    \includegraphics[width=\textwidth]{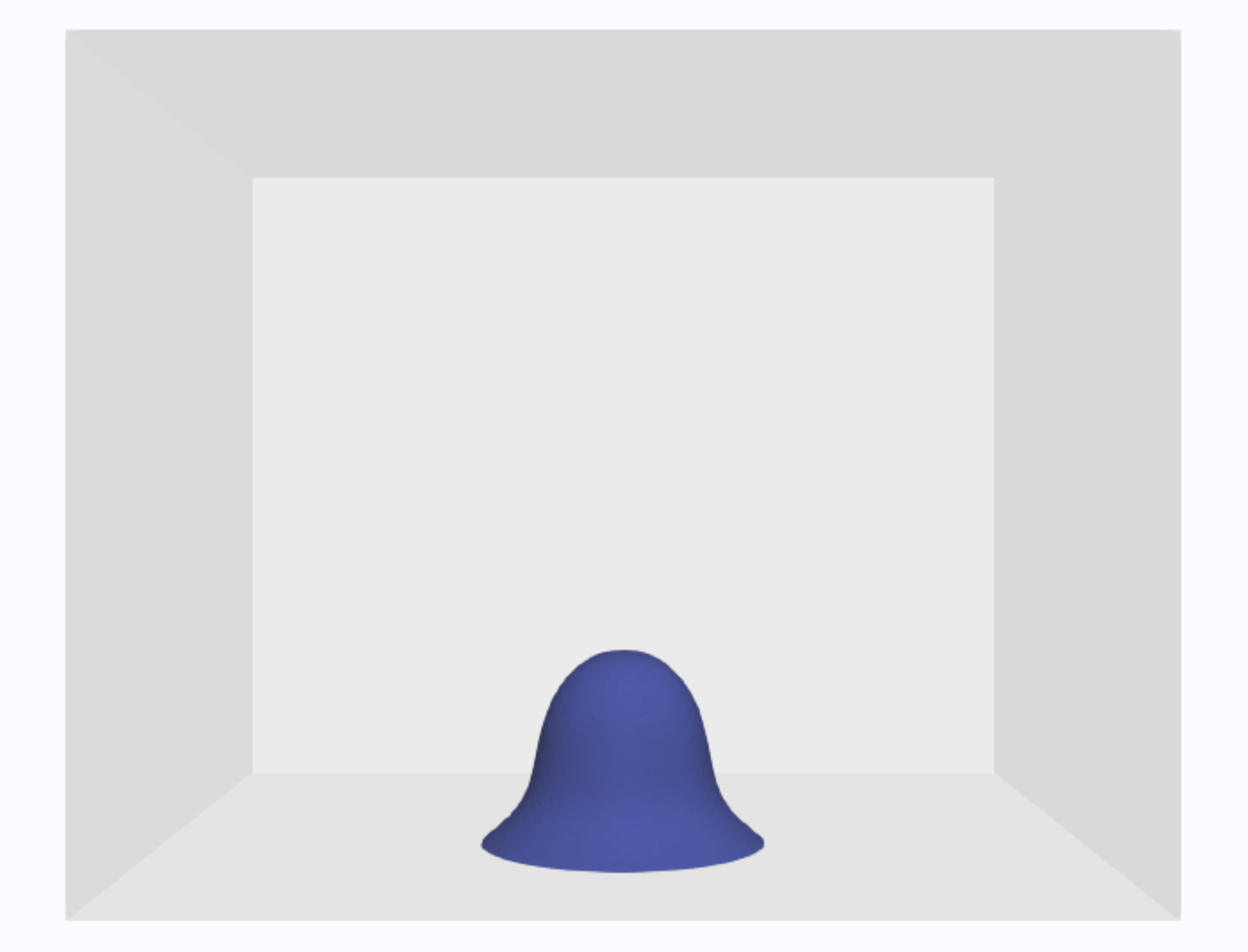}
    \caption{Time evolution, $t=0.5$}
    \label{fig:second}
\end{subfigure}
\hfill
\begin{subfigure}{0.4\textwidth}
    \includegraphics[width=\textwidth]{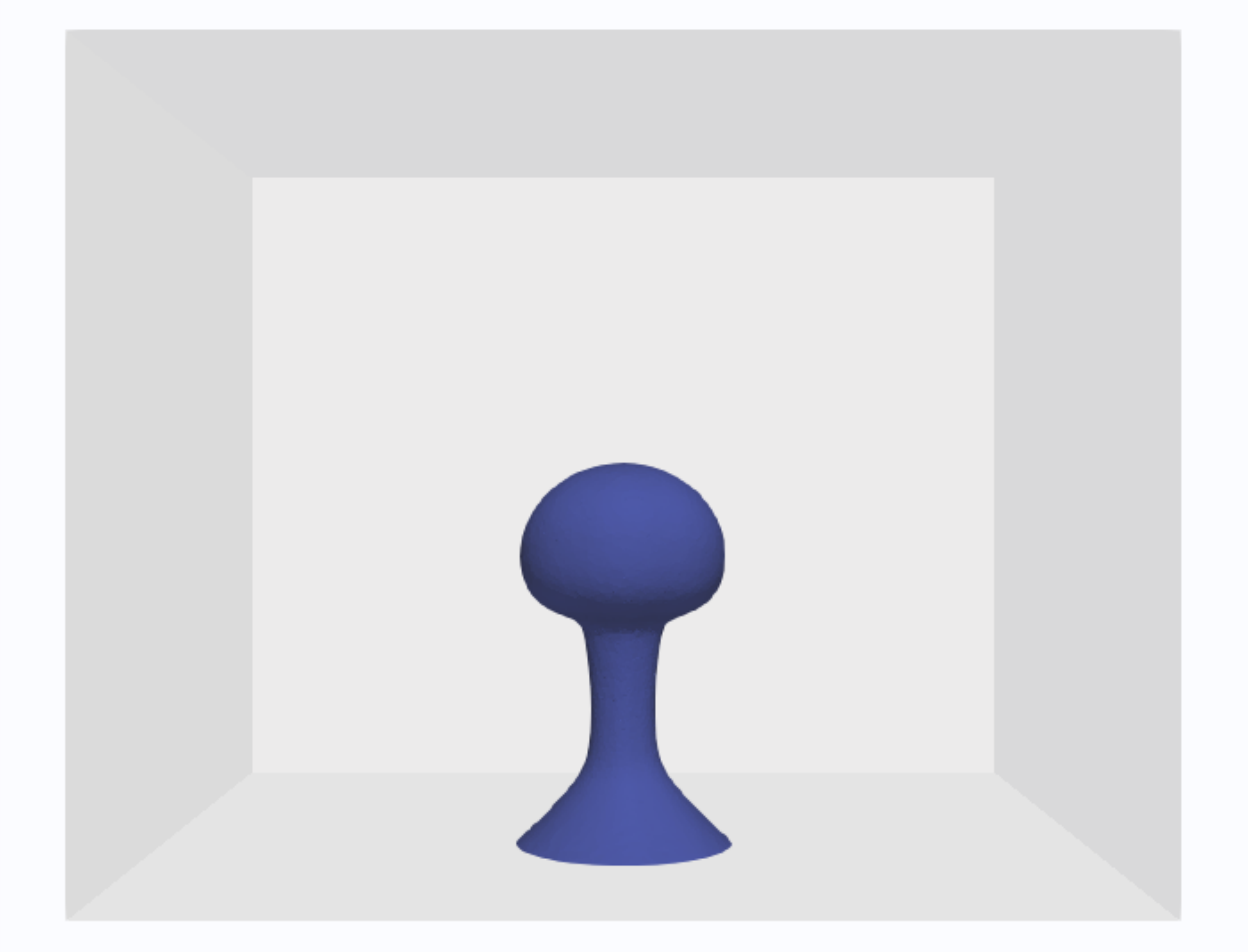}
    \caption{The detachment process, $t=1.2$.}
    \label{fig:third}
\end{subfigure}
\hfill
\begin{subfigure}{0.4\textwidth}
    \includegraphics[width=\textwidth]{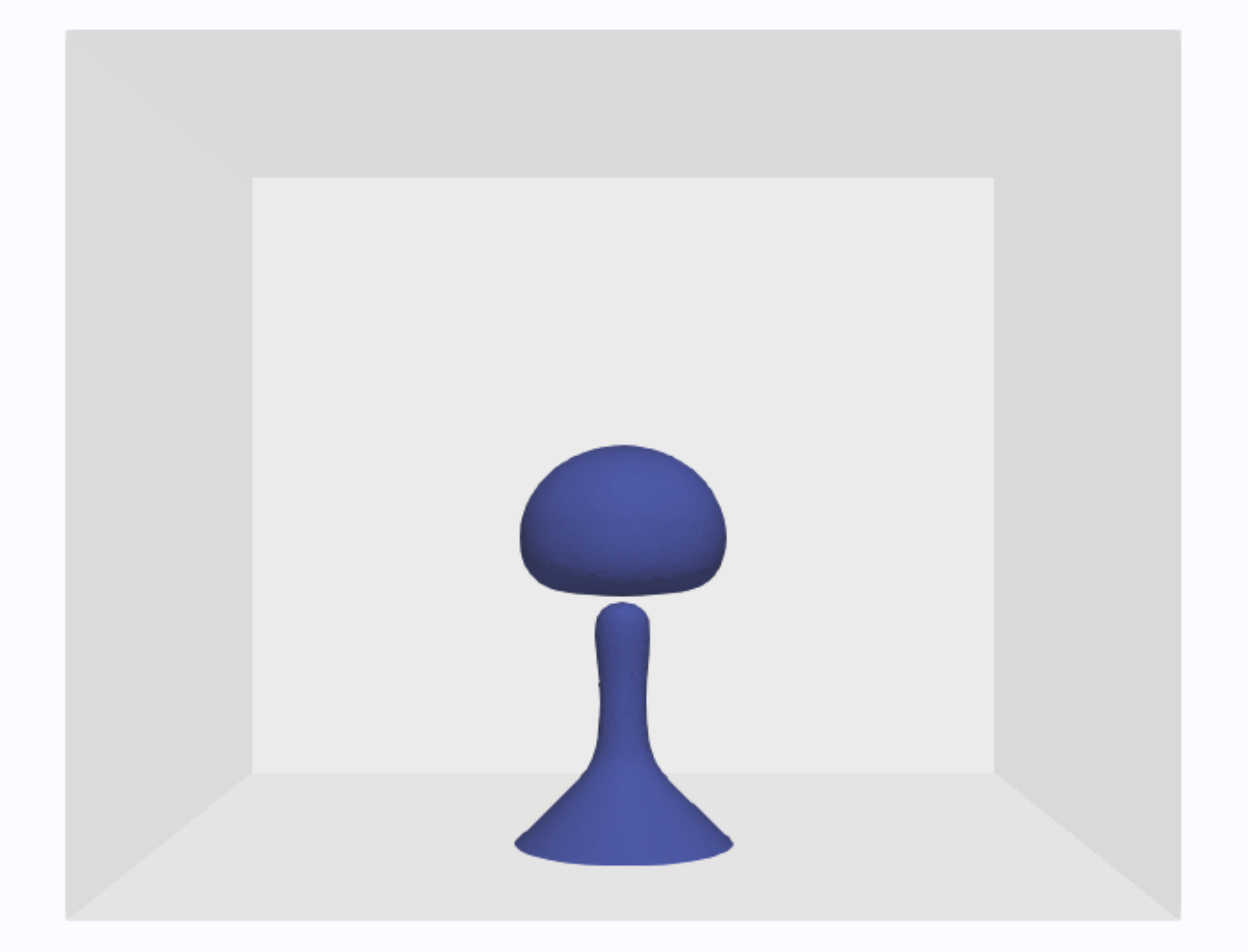}
    \caption{The break-up, $t=1.25$.}
    \label{fig:fourth}
\end{subfigure}
\hfill
\begin{subfigure}{0.4\textwidth}
    \includegraphics[width=\textwidth]{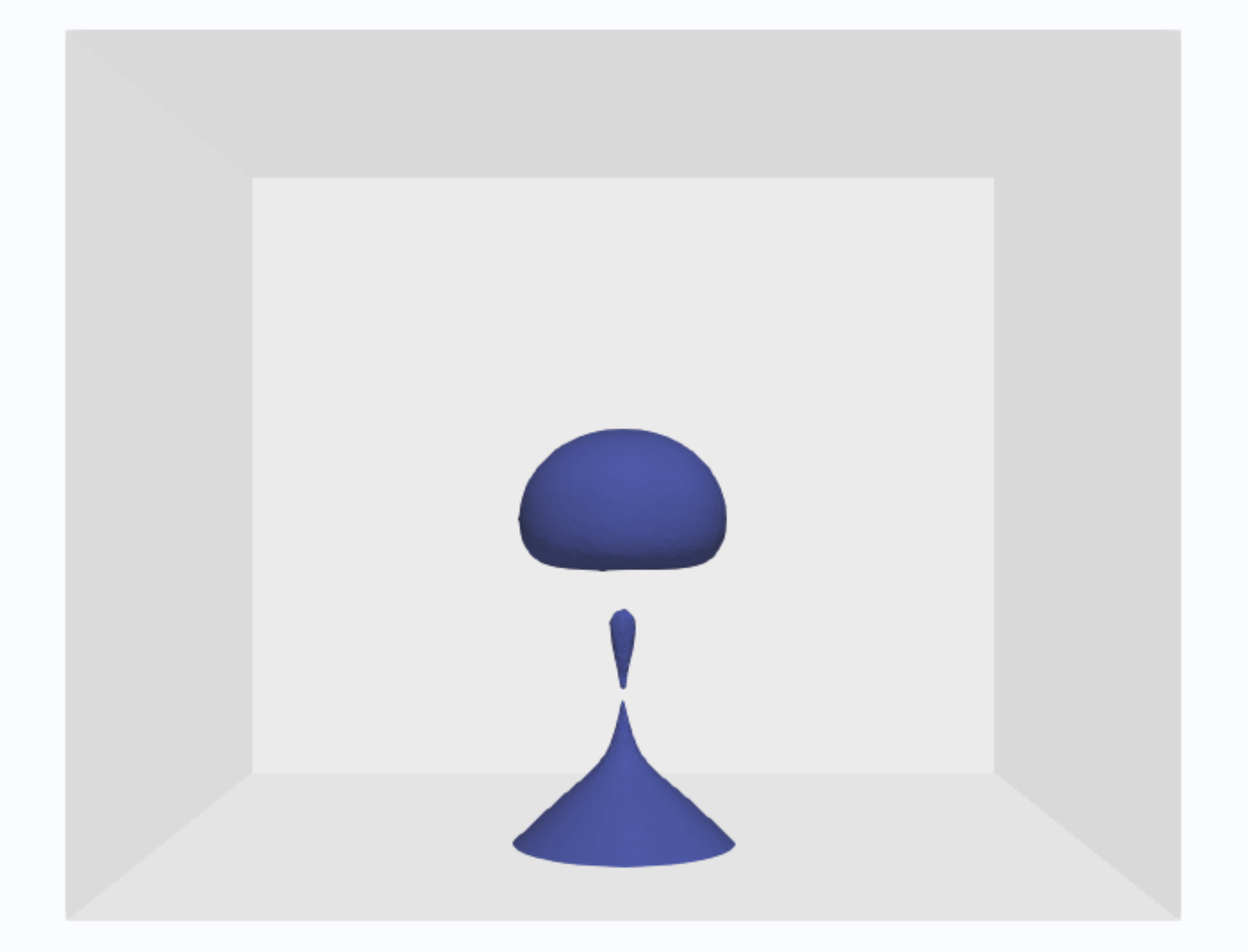}
    \caption{The formation of a satellite bubble, $t=1.3$.}
    \label{fig:fifth}
\end{subfigure}
\hfill
\begin{subfigure}{0.4\textwidth}
    \includegraphics[width=\textwidth]{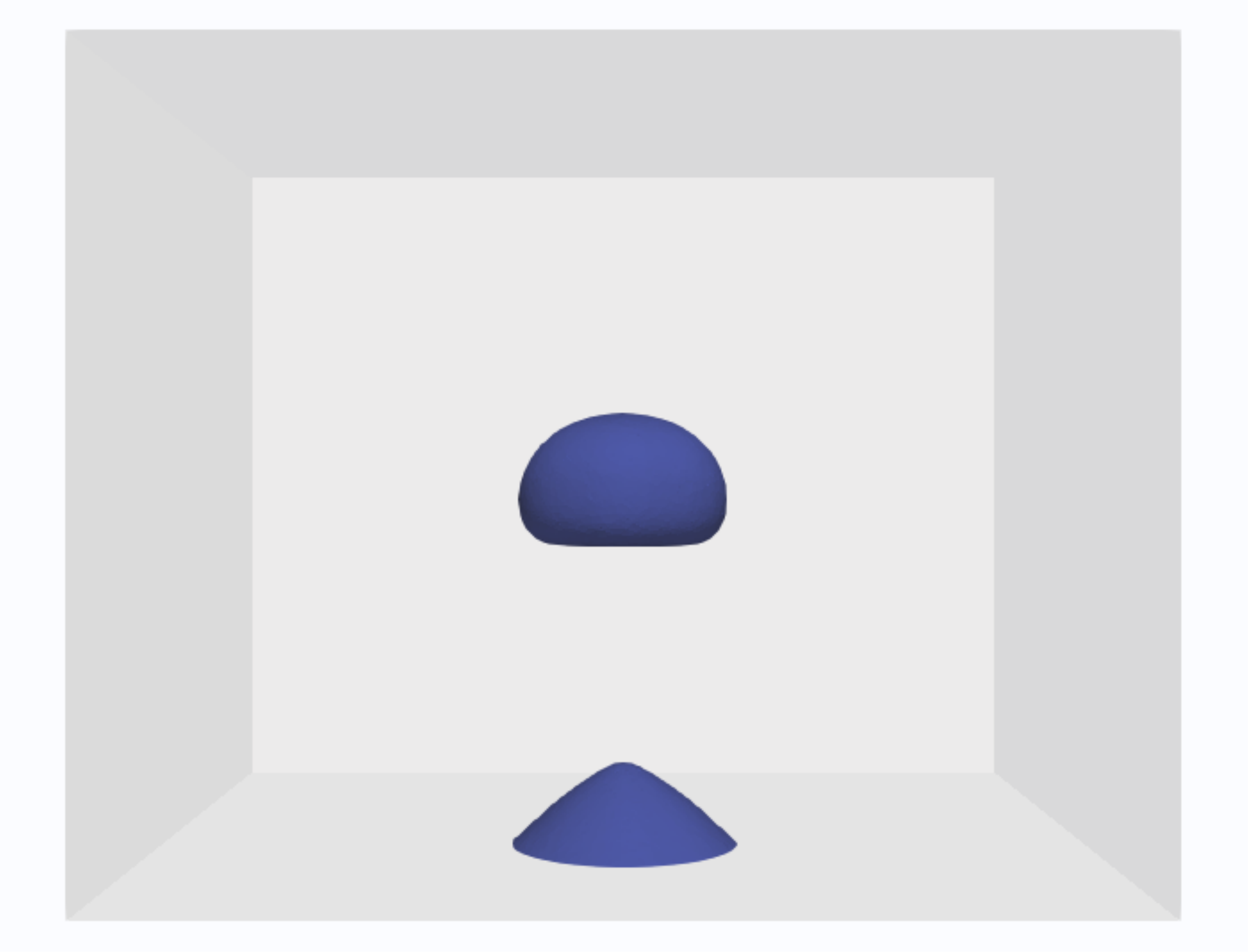}
    \caption{After the whole process part of the initial mass remains attached, $t=1.35$.}
    \label{fig:sixth}
\end{subfigure}
\caption{Numerical simulations of the evolution of a bubble with initial
contact angle $3\protect\pi/5$ and Bond number 15 (time in dimensionless
units).}
\label{fig3}
\end{figure}

In figure \ref{fig3} we plot some snapshots of the evolution of a bubble. We
have simulated the case $Bo=15$, and contact angle $\theta =3\pi /5$. At
those values, we are above the critical curve so we are through the process
of destabilization and detachment. It is noteworthy that bubbles completely
detach when the contact angle is sufficiently small, but for contact angles
closer to $\pi $ the detached bubble does not carry all the gas and part of
it is left at the substrate. Moreover, satellite bubbles do appear, much as
in the case of bubbles surrounded by inviscid liquids (cf. \cite{FG}). We
indeed observe this fact in the numerical simulations. We also observe that
the geometry of the interface near the pinch-off point where the bubble
breaks is symmetrical and close two two cones as described, for instance, in
(\cite{EF}).

% \begin{figure}[t]
% \center
% \includegraphics[width=0.7\textwidth]{bond15_3pi5_00}
% \includegraphics[width=0.7\textwidth]{bond15_3pi5_05}
% \includegraphics[width=0.7\textwidth]{bond15_3pi5_05}
% \includegraphics[width=0.7\textwidth]{bond15_3pi5_05}
% \includegraphics[width=0.7\textwidth]{bond15_3pi5_05}
% \includegraphics[width=0.7\textwidth]{bond15_3pi5_05}
% \caption{Bond critical number versus contact angle. Squares are numerical simularions, blue is theory and discontinuos line is the parabolic fitting of the numerical simulations.}
% \label{fig3}
% \end{figure}

\section{Inclined substrates}

\begin{figure}[t]
\center
\includegraphics[width=0.7\textwidth]{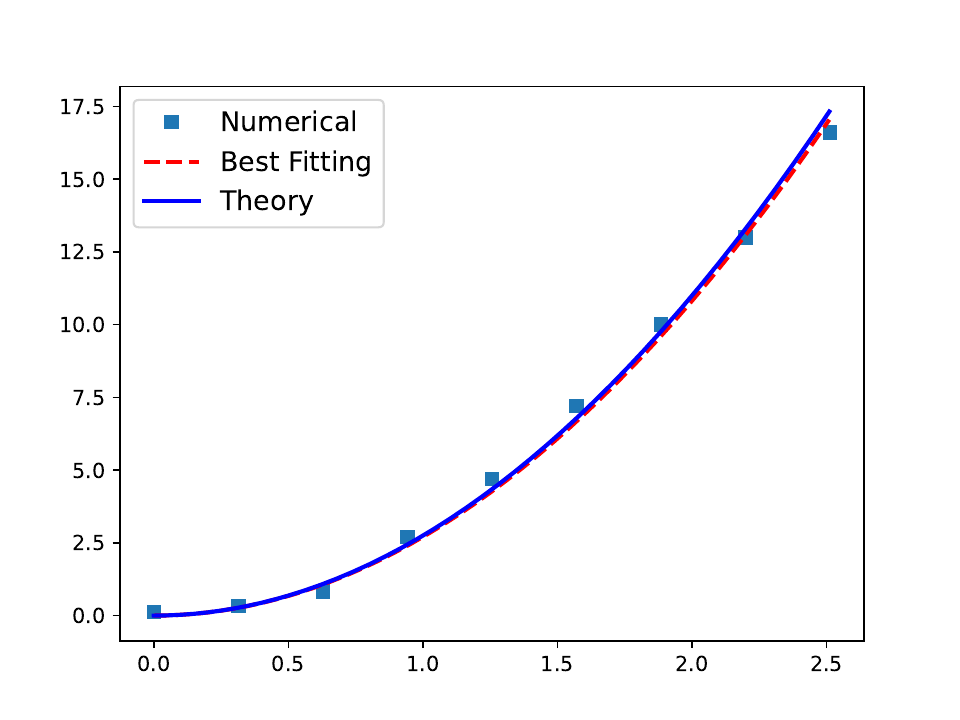}
\caption{Bond critical number versus contact angle for an inclined substrate
$\protect\alpha=30^\circ$. Squares are numerical simulations, blue is theory
and discontinuos line is the parabolic fitting of the numerical simulations $%
Bo=l\protect\theta^2$, being $l_{fit}=2.71$ and $l_{th}=2.38$.}
\label{fig4}
\end{figure}

We consider in this section the effect of an inclination angle $\alpha $ of
the substrate with respect to a substrate orthogonal to gravity. The
critical Bond number increases as we can see in figure \ref{fig4}. In fact,
for $\alpha =30%
%TCIMACRO{\U{ba}}%
%BeginExpansion
{{}^o}%
%EndExpansion
$ the critical Bond number is roughly the one corresponding to an
``effective" gravity given by $g\cos \alpha $. The effect is also apparent
for larger inclination angles as we can see in table \ref{table1} for different inclination angles. Again, a good approximation for the critical Bond number is the one
corresponding to an effective gravity $g\cos \alpha$ but a bit
larger. We can understand this as the effective combination of a new (and
smaller) effective gravity together with a deformation of the contact
gas/solid surface, which is no longer a circle and has a larger area,
leading to a higher resistence to detachment.

\begin{table}[htbp]
	\centering
	\renewcommand{\arraystretch}{1.3} % Ajustar el espaciado vertical
	\begin{tabular}{cccc}
		\toprule
		Angle & $B_o$ (No slope) & $B_o$ (inclined plane 30$^\circ$) & $B_o$ (inclined plane 60$^\circ$) \\
		\midrule
		$\pi/5$   & 0.7  & 0.8  & 1.35 \\
		$3\pi/10$ & 2.2  & 2.7  & 4.3 \\
		$2\pi/5$  & 4.0  & 4.7  & 8.2 \\
		$\pi/2$   & 5.9  & 7.2  & 12.0 \\
		$3\pi/5$  & 8.3  & 10.0 & 16.8 \\
		$7\pi/10$ & 10.8 & 13.0 & 22.0 \\
		$4\pi/5$  & 13.8 & 16.6 & 30.0 \\
		\bottomrule
	\end{tabular}
	\caption{Critical $B_o$ with different plane slopes}
	\label{table1}
\end{table}

\begin{figure}[t]
\center
\includegraphics[width=0.7\textwidth]{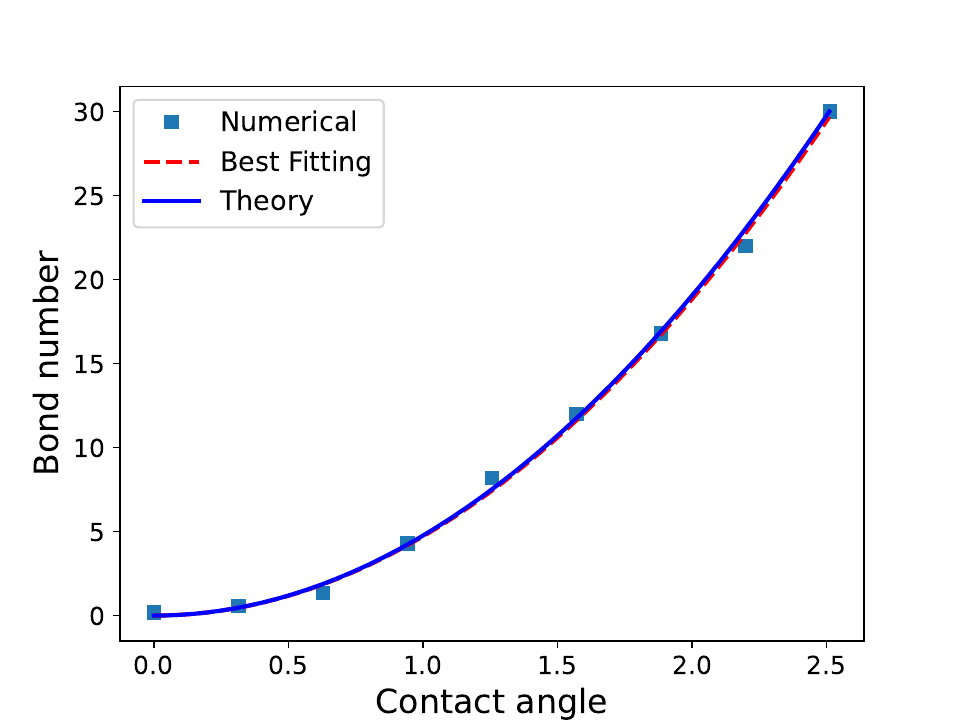}
\caption{Bond critical number versus contact angle for an inclined substrate
$\protect\alpha=60^\circ$. Squares are numerical simulations, blue is theory
and discontinuos line is the parabolic fitting of the numerical simulations $%
Bo=l\protect\theta^2$, being $l_{fit}=4.71$ and $l_{th}=2.38$.}
\label{fig5}
\end{figure}

In figure \ref{fig5} we can see that the critical Bond number increases as
compare with the theoretical value and follows the theoretical curve with an
effective gravity corrected by the inclination angle.

\section{The effect of external flows}

In this section we consider the bubble subjected to a external linear flow
of the form%
\begin{equation*}
\mathbf{u}_{in}=W_{in}z\, \mathbf{e}_{z},
\end{equation*}%
imposed far away from the bubble. We can introduce a characteristic length $%
R $ and velocity $U$, as well as the dimensionless number%
\begin{equation*}
w_{in}=\frac{W_{in}R}{U}=\frac{W_{in}R}{\sqrt{\rho _{1}gR}}\, ,
\end{equation*}%
in terms of which the condition of infinity reads%
\begin{equation*}
\mathbf{u}_{in}=w_{in}z\, \mathbf{e}_{z}.
\end{equation*}%
Therefore, we have the following set of dimensionless numbers:%
\begin{equation*}
Bo,Re,w_{in}.
\end{equation*}

\bigskip If we include kinetic energy into the energetic balance (with an
undetermined prefactor) we arrive at the relation%
\begin{equation*}
Bo=\frac{(\rho _{L}-\rho _{G})gR^{2}}{\sigma _{LG}}=\frac{3}{2^{\frac{1}{3}}}%
(1-\cos \theta )-\delta w_{in}^{2},
\end{equation*}%
which is a parabolic dependence on $w_{in}^{2}$.

\begin{figure}[t]
\center
\includegraphics[width=0.7\textwidth]{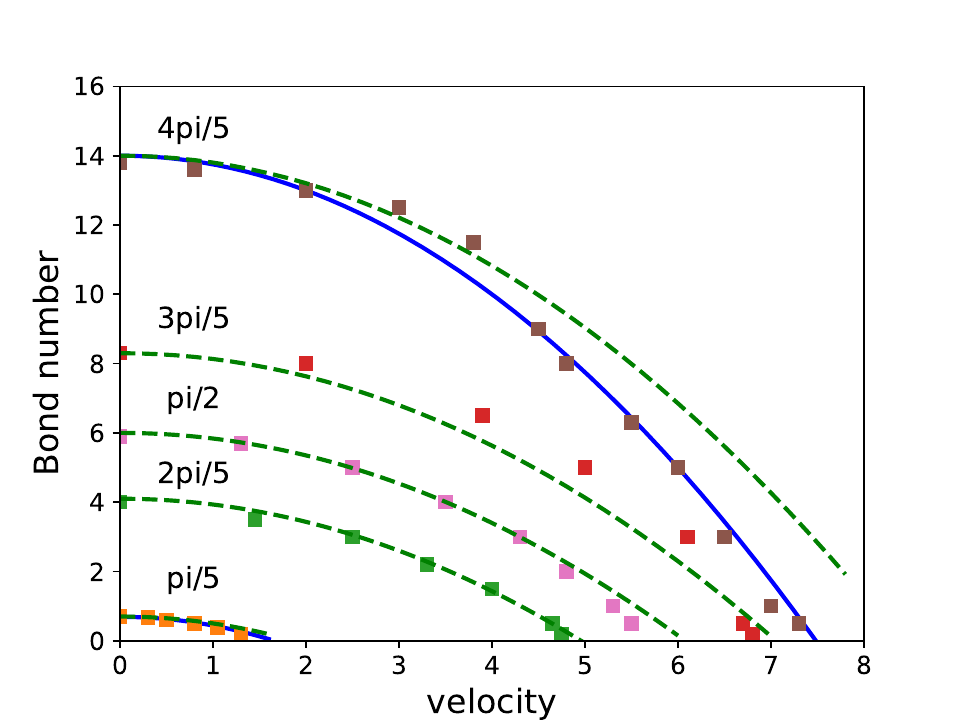}
\caption{Bond critical number versus inlet velocities at different contact
angles. Squares are numerical simulations, discontinuos lines are given by $%
Bo=l_{th}\protect\theta ^{2}-\protect\delta \,w_{in}^{2}$ being $l_{th}=2.38$
and $\protect\delta =0.25(1-0.35\sin \protect\theta )$. The continuous line
is the one with $\protect\delta =0.25$. For small angles ($\protect\theta %
=\pi/5$) there is not difference between discontinuous and continuous curves.}
\label{fig6}
\end{figure}

Our simulations (see figure \ref{fig6}) indicate that $Bo$ indeed depend
cuadratically on $w_{in}^{2}$ with $\delta $ depending mildly on $\theta $.
In fact, for both small and large values of $\theta $ we found $\delta
\simeq 0.25$. The dashed curves correspond to the asymtotic expression
\begin{equation*}
Bo=\frac{3}{2^{\frac{1}{3}}}\theta^2-\frac{1}{4}(1-0.35\sin{\theta)}
w_{in}^{2}  \label{BoVin}\, .
\end{equation*}

The critical surface which separates the attaching-dettaching behaviour,
given by
\begin{equation}
Bo-3/2^{1/3}\theta ^{2}+0.25(1-0.35\sin \theta )w_{in}^{2}=0,
\label{critical}
\end{equation}%
together with the numerical simulation points are depicted in figure \ref%
{fig3d}. As we can see in the figure, the agreement of formula (\ref%
{critical}) with the numerical results is excellent except when $\theta $ is
close to $\pi $ for large velocities. The reason for this is that in that
case the velocity field is able to blow a portion of the bubble before the
contact line has a significant dynamics and a better approximations is given
by replacing the factor $\frac{1}{4}(1-0.35\sin \theta )$ by $\frac{1}{4}$,
which is independent of the contact angle $\theta .$

\begin{figure}[t]
\center
\includegraphics[width=1.0\textwidth]{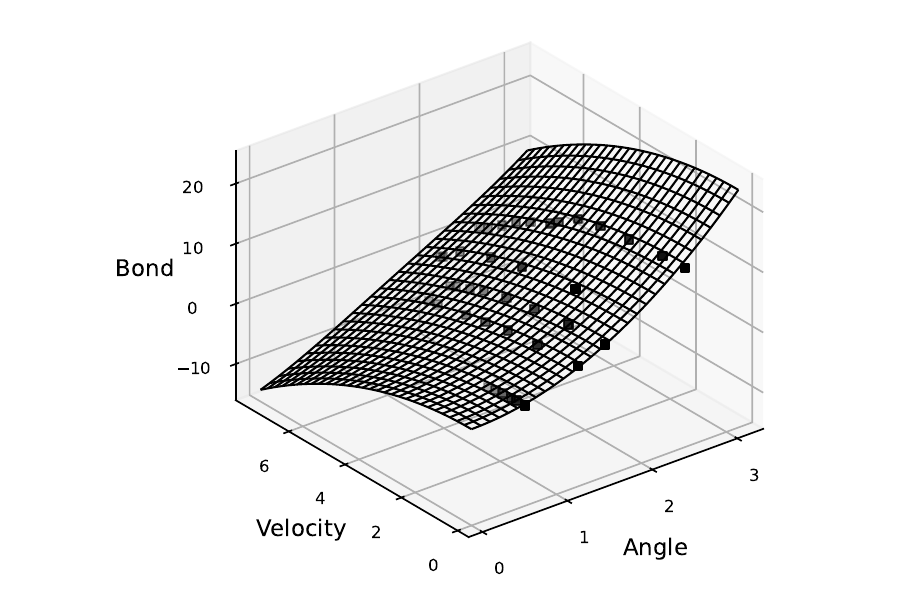}
\caption{Critical surface. Squares points are numercial simulations,
critical surface is given by $Bo-l_{th}\protect\theta^2 + \protect\delta\,
w_{in}^{2}=0$, being $l_{th}=3/2^{1/3}$ and $\protect\delta=0.25(1-0.35\sin%
\protect\theta)$.}
\label{fig3d}
\end{figure}

\begin{figure}[t]
\centering
\begin{subfigure}{0.45\textwidth}
    \includegraphics[width=\textwidth]{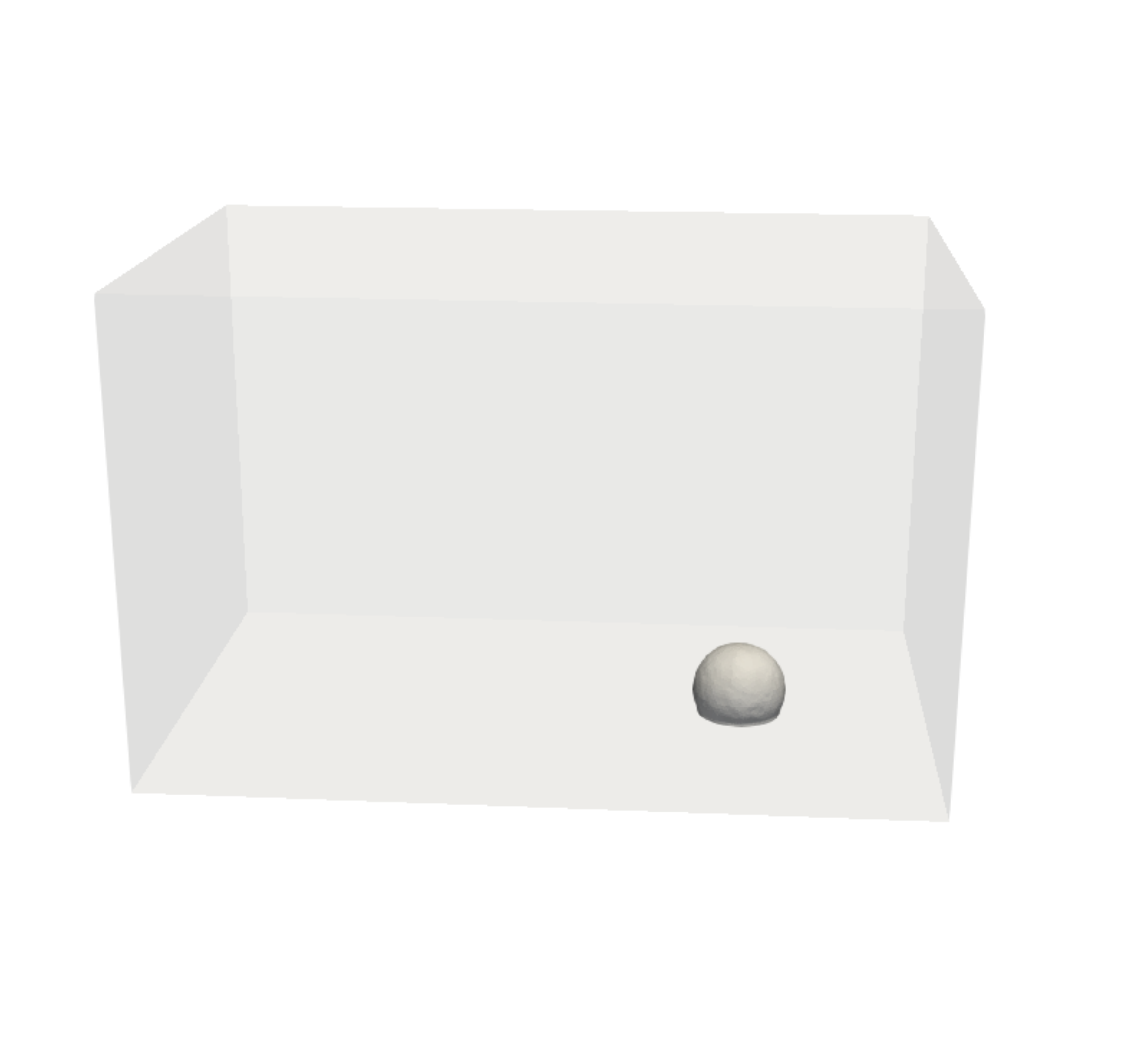}
    \caption{Initial state, $t=0$.}
    \label{fig2:first_in}
\end{subfigure}
\hfill
\begin{subfigure}{0.45\textwidth}
    \includegraphics[width=\textwidth]{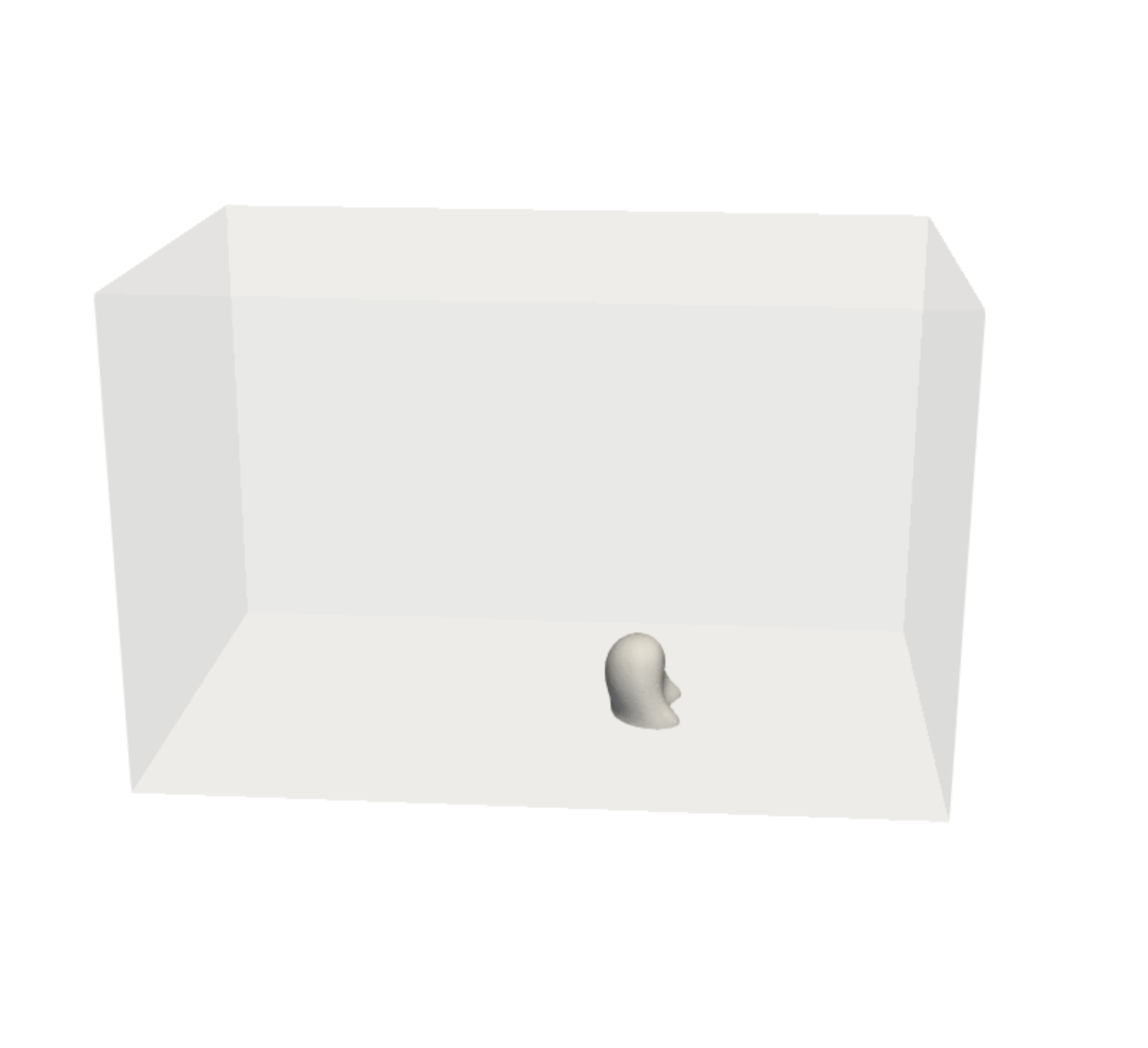}
    \caption{Evolution under inflow, $t=0.2$}
    \label{fig2:second_in}
\end{subfigure}
\hfill
\begin{subfigure}{0.45\textwidth}
    \includegraphics[width=\textwidth]{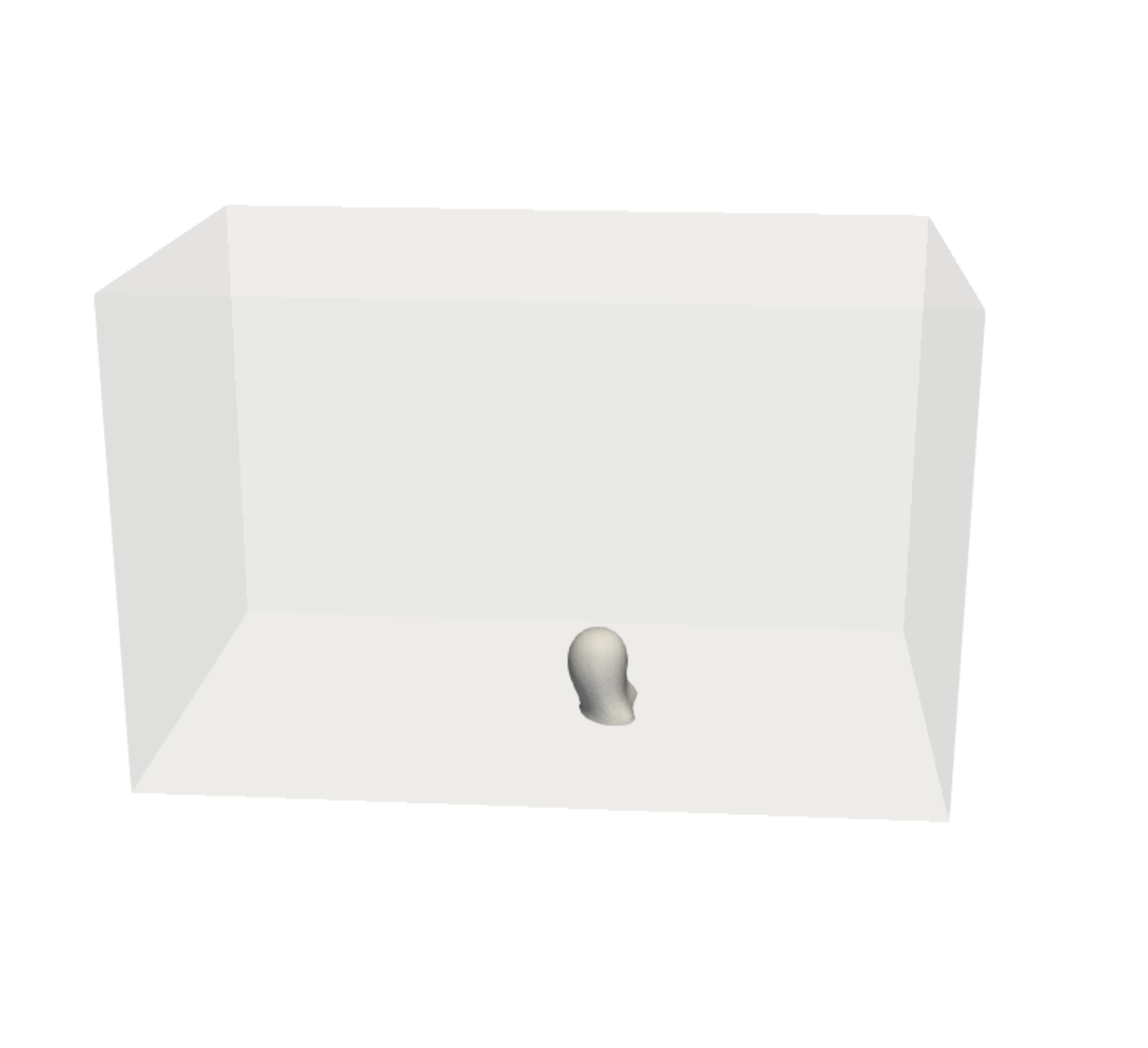}
    \caption{Time evolution, $t=0.3$.}
    \label{fig2:third_in}
\end{subfigure}
\hfill
\begin{subfigure}{0.45\textwidth}
    \includegraphics[width=\textwidth]{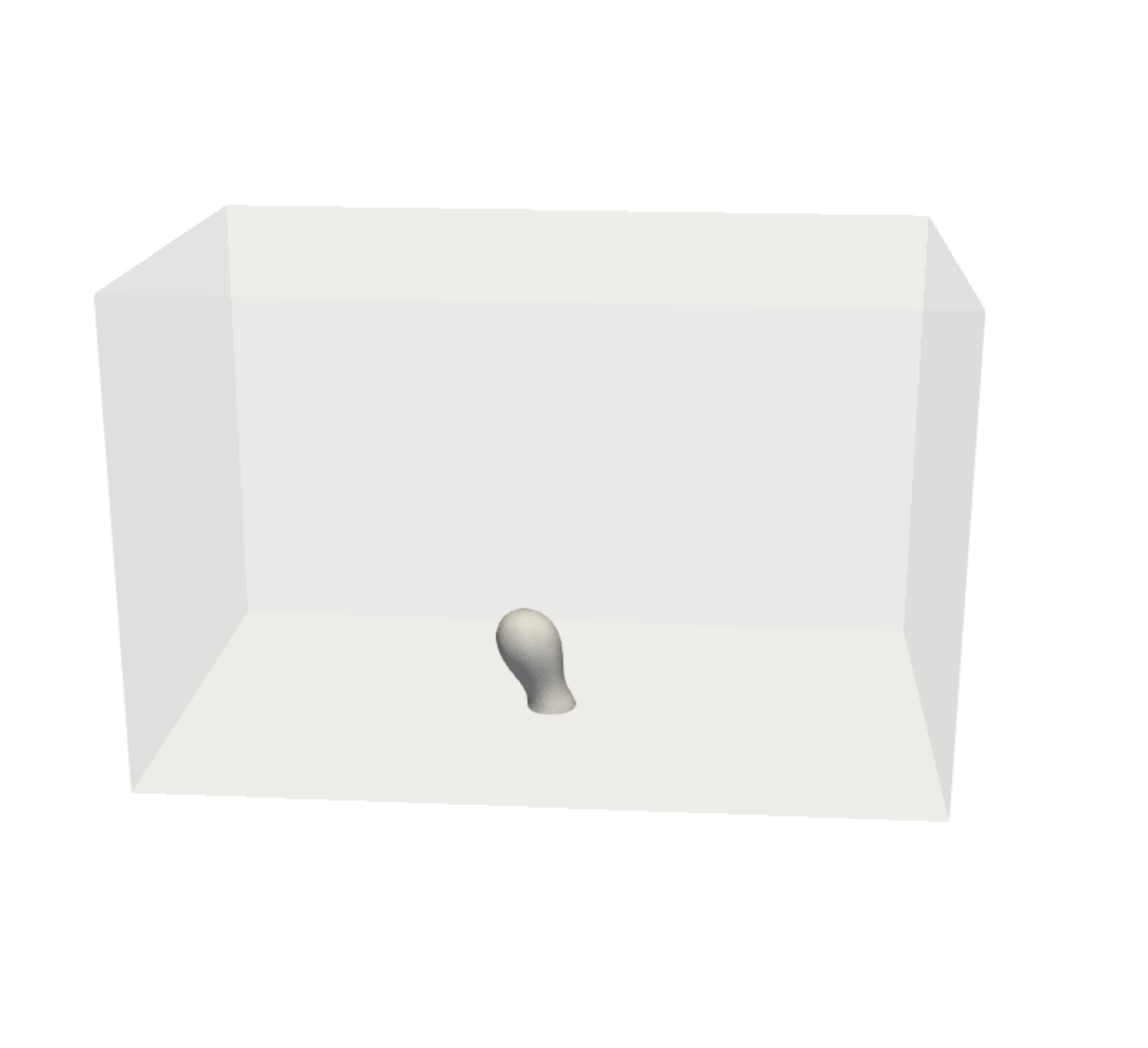}
    \caption{The detachment process, $t=0.5$.}
    \label{fig2:fourth_in}
\end{subfigure}
\hfill
\begin{subfigure}{0.45\textwidth}
    \includegraphics[width=\textwidth]{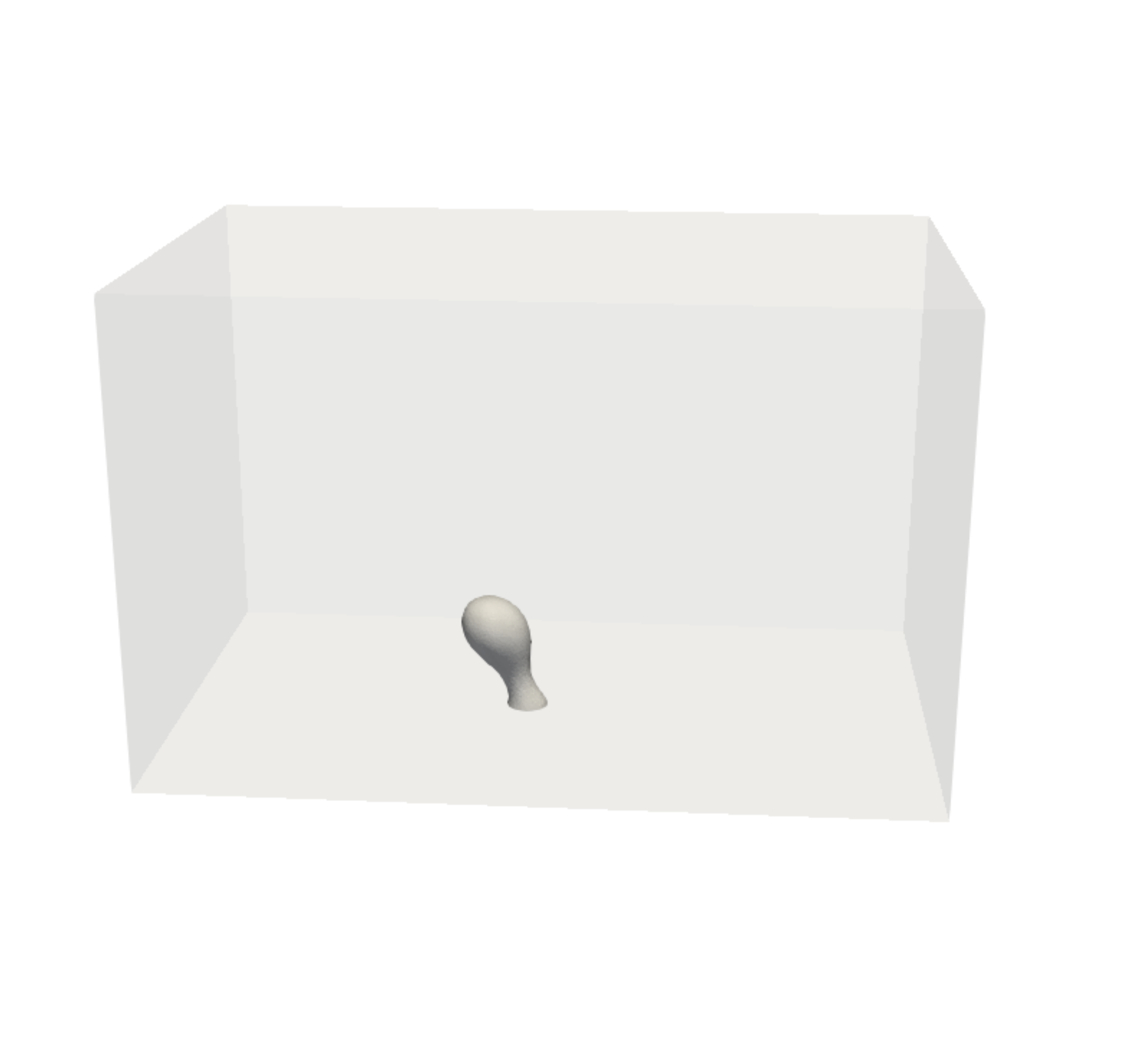}
    \caption{Formation of a neck, $t=0.6$.}
    \label{fig2:fifth_in}
\end{subfigure}
\hfill
\begin{subfigure}{0.45\textwidth}
    \includegraphics[width=\textwidth]{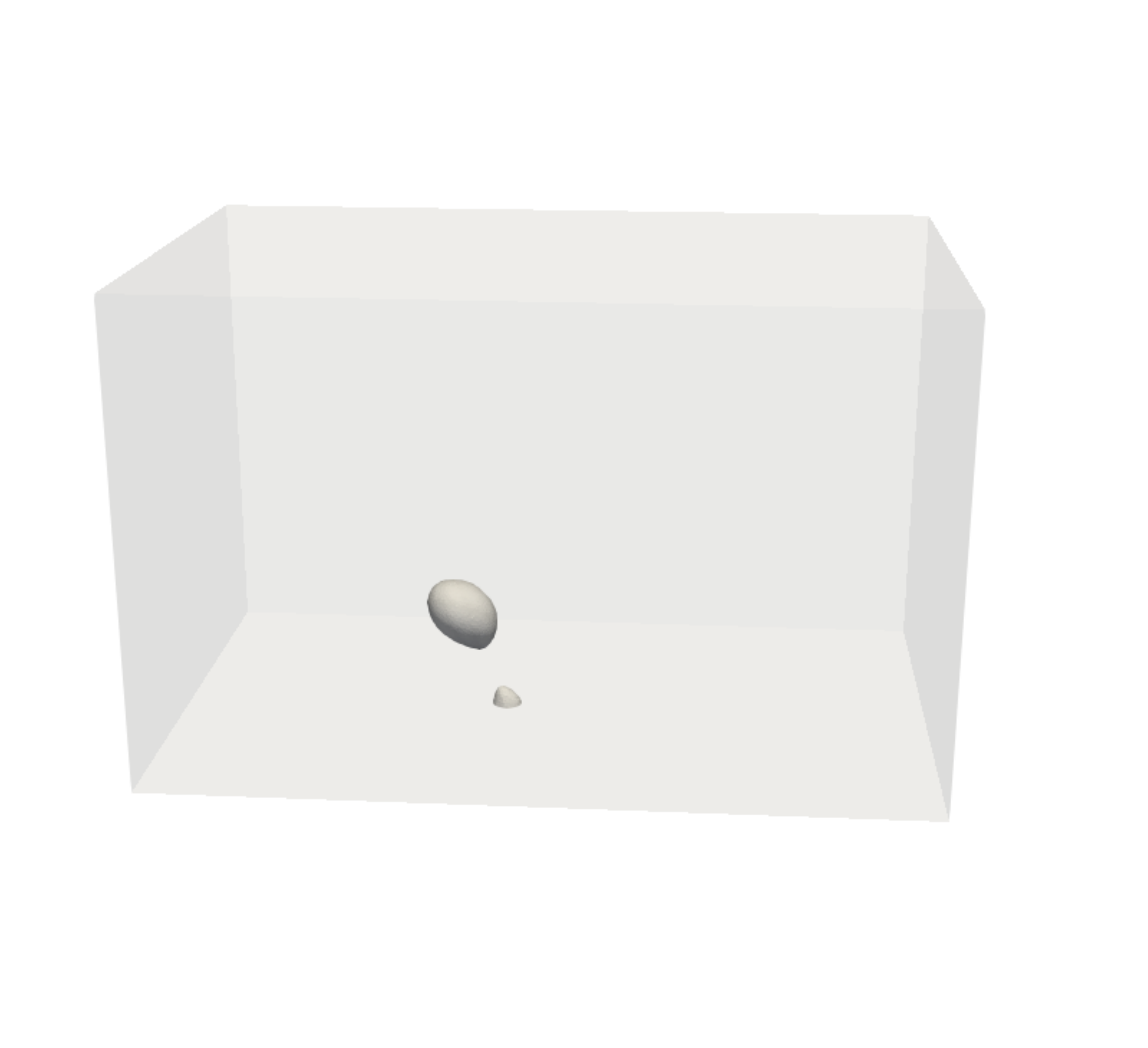}
    \caption{The break-up, $t=0.7$.}
    \label{fig:sixth_in}
\end{subfigure}
\caption{Numerical simulations of the evolution of a bubble with initial
contact angle $3\protect\pi /5$, Bond number 10 (time in dimensionless
units). There is an inlet linear flow $v_{y}=-Uz$ with $U=3$ in
dimensionless units.}
\label{fig7}
\end{figure}

In figure \ref{fig7} we plot various stages of the evolution of the bubble
under the action of an external velocity field as computed by the phase
field method. The Bond number is 15 and the initial contact angle is $3\pi/5$%
. The inlet velocity is a laminar flow in the right yx-plane, given by $%
v=-Uz $, with $U=3$.

\section{Conclusions}

We have developed and implemented a phase field model for the study of the
evolution and detachment from a solid substrate of a gas bubble surrounded
by a viscous liquid. Three different physical conditons have been
considered: 1) bubble surrounded by a quiescent viscous fluid and under the
action of a vertical gravitational field, 2) bubble over an inclined plane,
3) bubble under the action of a fluid flow. We found that, in terms of
dimensionless numbers, the condition for bubble detachment follows very
simple laws with a strong dependence on the static Young's contact angle.
These laws are quadratic with a high degree of accuracy. As a side result of
our study, the phase field model is able to handle topological changes
related to the pinch-off of the bubble and the formation of satellites.

If future publications we will study the detachment of bubbles that grow due
to the diffusion of a gas (typically hydrogen or oxygen) inside the bubble.
The gas will be produced on the solid substrate due to chemical reactions
such as those associated to hydrogen production in water electrolysis. Our
goal is to analyse the interplay between bubble growth and detachment and
effective electrode area for chemical reaction in order to optimize the
conditions for hydrogen production. In this sense, the present work helps to
understand how physical parameters and mechanisms such as contact angle and
external flow influence the production of gas.

\begin{acknowledgement}
This work has been supported by projects TED2021-131530B-I00 and
PID2022-139524NB-I00.
\end{acknowledgement}

\newpage

% \begin{figure}[b]
% \includegraphics[width=0.15\textwidth]{figure_1.pdf}
% \caption{Critical Bond number vs contact angle.}
% \label{fig2}
% \end{figure}

% \newpage

% \begin{figure}[b]
% \includegraphics[width=0.15\textwidth]{image.png}
% \caption{Critical Bond number vs contact angle.}
% \label{fig3}
% \end{figure}

% \newpage

% \begin{figure}[b]
% \includegraphics[width=0.15\textwidth]{figure_2.pdf}
% \caption{Critical Bond number vs dimensionless velocity for various contact
% angle.}
% \label{fig5}
% \end{figure}

\end{document}